\documentclass[doublespacing]{bmcart}

\usepackage{amsthm,amsmath}
\usepackage[utf8]{inputenc} 

\usepackage[clean]{revdiff}
\usepackage{cite}
\usepackage{amsmath,amssymb,amsfonts}
\usepackage{algorithmic}
\usepackage{graphicx}
\usepackage{textcomp}

\usepackage{anyfontsize}
\usepackage[resetcount,lined,boxed,commentsnumbered,ruled,vlined,longend]{algorithm2e}
\newcommand{\nextnr}{\stepcounter{AlgoLine}}
\makeatletter
\newcommand{\removelatexerror}{\let\@latex@error\@gobble}
\makeatother
\usepackage{array}
\usepackage[caption=false,font=normalsize,labelfont=sf,textfont=sf]{subfig}
\usepackage{fixltx2e}
\usepackage{stfloats}
\usepackage{setspace}

\let\MYorigsubfloat\subfloat
\renewcommand{\subfloat}[2][\relax]{\MYorigsubfloat[]{#2}}
\usepackage{url}
\usepackage{amssymb}

\usepackage{soul}
\usepackage{xcolor}
\usepackage{lineno}

\def\L{{\cal L}}

\def\r{\boldsymbol{r}}
\def\z{\boldsymbol{z}}

\def\E{{\mathcal{E}}}

\def\A{\mathcal{A}}
\def\bA{\mathbf{A}}
\def\bAhat{\mathbf{\hat{A}}}

\def\cmtr{\mathrm{cm}}
\def\etal{\textit{et al.\ }}
\def\th{\mathrm{th}}

\def \tht{\boldsymbol{\theta}}

\def \attn {\boldsymbol{\mu}}
\def \obj {\boldsymbol{\lambda}}

\def \L {\mathcal{L}}

\def \P {\mathbb{P}}
\def \path{\mathbb{P}}

\def \E{E}
\def \f{\boldsymbol{\lambda}}

\def \subpath{\mathbb{S}}
\def \pr{\mathrm{pr}}
\def \Pr{\mathrm{Pr}}

\def \ra_223{^{223}\text{Ra}}

\startlocaldefs
\endlocaldefs

\begin{document}

\begin{singlespace}
\begin{frontmatter}

\begin{fmbox}
\rlegend

\begin{huge}\textbf{A list-mode multi-energy window low-count SPECT reconstruction method for isotopes with multiple emission peaks}\end{huge}

\author[
  addressref={aff1,aff2},                   
  email={rahman.m@wustl.edu}   
]{\inits{M.A.R}\fnm{Md Ashequr} \snm{Rahman}}
\author[
  addressref={aff1,aff2},
  email={zekunli@wustl.edu}
]{\inits{Z.L.}\fnm{Zekun} \snm{Li}}
\author[
  addressref={aff1,aff2},
  email={yu.zitong@wustl.edu}
]{\inits{Z.Y.}\fnm{Zitong} \snm{Yu}}
\author[
  addressref={aff2},
  email={rlaforest@wustl.edu}
]{\inits{R.L.}\fnm{Richard} \snm{Laforest}}
\author[
  addressref={aff2},
  email={thorekd@wustl.edu}
]{\inits{D.T.}\fnm{Daniel L.J.} \snm{Thorek}}
\author[
  addressref={aff1,aff2},
  corref={aff1,aff2}, 
  email={a.jha@wustl.edu}
]{\inits{A.K.J.}\fnm{Abhinav K.} \snm{Jha}}

\address[id=aff1]{
  \orgdiv{Department of Biomedical Engineering},             
  \orgname{Washington University in St. Louis},          
  \city{St. Louis},                              
  \cny{USA}                                    
}
\address[id=aff2]{
  \orgdiv{Mallinckrodt Institute of Radiology},             
  \orgname{Washington University in St. Louis},          
  \city{St. Louis},                              
  \cny{USA}                                    
}



\end{fmbox}

\begin{abstract} 
\parttitle{Background} 
Single-photon emission computed tomography (SPECT) provides a mechanism to perform absorbed-dose quantification tasks for $\alpha$-particle radiopharmaceutical therapies ($\alpha$-RPTs). However, quantitative SPECT for $\alpha$-RPT is challenging due to the low number of detected counts, the complex emission spectrum, and other image-degrading artifacts. Towards addressing these challenges, we propose a low-count quantitative SPECT reconstruction method for isotopes with multiple emission peaks.

\parttitle{Methods}
Given the low-count setting, it is important that the reconstruction method extract the maximal possible information from each detected photon. Processing data over multiple energy windows and in list-mode (LM) format provide mechanisms to achieve that objective. Towards this goal, we propose a list-mode multi-energy window (LM-MEW) ordered-subsets expectation-maximization-based SPECT reconstruction method that uses data from multiple energy windows in LM format, and includes the energy attribute of each detected photon. For computational efficiency, we developed a multi-GPU-based implementation of this method. The method was evaluated using 2-D SPECT simulation studies in a single-scatter setting conducted in the context of imaging [$^{223}$Ra]RaCl${_2}$, an FDA-approved RPT for metastatic prostate cancer.

\parttitle{Results} 
The proposed method yielded improved performance on the task of estimating activity uptake within known regions of interest in comparison to approaches that use a single energy window or use binned data. The improved performance was observed in terms of both accuracy and precision and for different sizes of the region of interest.
\parttitle{Conclusions} 
Results of our studies show that the use of multiple energy windows and processing data in LM format with the proposed LM-MEW method led to improved quantification performance in low-count SPECT of isotopes with multiple emission peaks. These results motivate further development and validation of the LM-MEW method for such imaging applications, including for $\alpha$-RPT SPECT.

\end{abstract}


\begin{keyword}
\kwd{Quantitative SPECT}
\kwd{Reconstruction}
\kwd{List-mode data}
\kwd{Theranostics}
\kwd{Objective task-based evaluation}
\kwd{$\alpha$-particle radiopharmaceutical therapies}
\kwd{Radium-223}
\end{keyword}


%

\end{frontmatter}
\end{singlespace}
\section{Background}
Radiopharmaceutical therapies (RPTs) with $\alpha$-particle emitting isotopes are showing significant promise in multiple clinical and pre-clinical studies \cite{abou2020prostate,baidoo2013molecular}. This cancer therapy approach sees the localized emission of cytotoxic, high linear energy transfer, helium nuclei at sites of disease. Several such therapies are being actively investigated or recently approved, including those based on radium-223 \cite{kluetz2014radium}, actinium-225 \cite{kratochwil2017targeted}, and thorium-227 \cite{larsson2020feasibility,murray2020quantitative,ghaly2019quantitative}. The $\alpha$-emitting RPTs ($\alpha$-RPTs) are designed to target tumors, but these isotopes can also be absorbed by other regions inside the body and can potentially damage normal tissues and vital organs \cite{gustafsson2020feasibility,tafreshi2019development}. Thus, it is essential to quantify the absorbed dose in both target lesions and in at-risk organs in the application of $\alpha$-RPTs. In addition, quantification of absorbed dose can help to adapt treatment regimens, predict therapy outcomes, and monitor adverse events \cite{brans2007clinical}.

Since the decay of $\alpha$-emitting radionuclides usually also emit photons, single-photon emission computed tomography (SPECT) provides a mechanism for \textit{in vivo} reconstruction of the isotope activity distribution and the resultant absorbed-dose distribution in $\alpha$-RPTs \cite{benabdallah2019223ra,osti_22632144,gustafsson2020feasibility,li2021projection,abou2020preclinical}. However, performing this reconstruction task is challenging. 
A key reason is that due to \rchange{the deposition of highly cytotoxic MeV doses along a length scale of only several cells by $\alpha$-RPTs}{high linear energy transfer of $\alpha$-particles}, \rnew{a }low amount of activity is typically \rchange{injected}{administered} – of the order of $\sim$10 MBq - to reach therapeutic effectiveness.
This results in the detected number of counts when performing SPECT for $\alpha$-RPTs to be very small, around three orders of magnitude lower than that in conventional quantitative SPECT (QSPECT) applications \cite{owaki2017ra,benabdallah2021practical}.
Another major reason is that the emission of $\alpha$-emitting isotopes typically follow a complex cascade of several $\alpha$ (and $\beta$) emissions through several daughter nuclides that themselves emit photons. Thus, the photon emission spectra of these radionuclides are complex.
A sample Monte Carlo (MC)-simulated energy spectrum is shown in Fig. \ref{fig:MC_spectra} for $^{223}$Ra-based $\alpha$-RPT SPECT. Additionally, the image-degrading effects in SPECT such as noise, attenuation, scatter, and collimator-detector response further complicate the quantification process.
It is observed that conventional QSPECT methods yield a high value of bias (19\%-35\%) and standard deviation (12\%-30\%) in the estimated activity uptake of different regions of interest \cite{benabdallah2019223ra,osti_22632144,gustafsson2020feasibility}. To address these challenges in $\alpha$-RPT SPECT, there is an important need to develop new low-count quantitative SPECT reconstruction methods for isotopes with multiple emission peaks.

Since the number of detected counts can be very small for applications including $\alpha$-RPT SPECT, for reliable quantification, it is imperative that the designed method process most of the detected photons and extract the maximal possible information from each detected photon.
In that context, SPECT systems acquire data over a wide energy spectrum, collect data in list-mode (LM) format, and, as part of that, also measure the energy attribute for each detected photon.
Each of these capabilities provides an opportunity to improve the performance of QSPECT methods at these low-count levels.
As stated earlier, $\alpha$-emitting isotopes follow a complex emission pattern. Thus, the emission spectra of $\alpha$-emitting isotopes usually consist of multiple photopeaks, and data can be collected over multiple energy windows corresponding to these photopeaks.
Processing data over these multiple windows enables extracting information from a larger number of photons. 
It was shown that using data from multiple energy windows can theoretically improve the precision in estimating uptake in regions of interest for Y-90 SPECT \cite{chun2019algorithms}.
In a singular value decomposition-based investigation, it was shown that modeling scatter in multiple energy windows improved the noise characteristics of reconstructed images \cite{kadrmas1996svd,kadrmas1997analysis}.
Further, with $\alpha$-particle SPECT, using multiple energy windows improves the detected counts by 40\% \cite{benabdallah2019223ra}, increasing the effective sensitivity of the system. 
Thus, using data from multiple energy windows may improve performance of QSPECT for $\alpha$-RPTs.

Next, processing the data in LM format provides a mechanism to extract the maximal information content from each photon. In LM data format, the attributes of each detected photon such as the position of the interaction in the scintillation detector, the energy deposited by the detected photon in the detector, and the time of detection can be recorded. Typically, these attributes are binned, which leads to loss of information. Since LM data does not suffer from binning-related information loss unlike the typical binned data, it provides an opportunity to extract maximal information from detected photons.
Previous studies have shown that processing the data in LM format, as opposed to the binned format, can improve performance \cite{clarkson2020quantifying,jha2015estimating,jha2015singular,caucci2019towards},
including on estimation tasks \cite{jha2015estimating,rahman2020ieee}.
More specifically, using the energy attribute of the detected LM event may improve quantification.
For example, using the energy information may improve scatter compensation. This is because the primary scatter mechanism as the photon traverses through human body is Compton scatter, and in this scatter process, the angle of scatter is directly related to the energy of the scattered photon \cite{bousse2016joint,arridge2021overview}.
Thus, the energy attribute constrains the potential paths that a scattered photon may have taken before being detected \cite{rahman2020fisher}.
In fact, Gu{\'e}rin \textit{et al} \cite{guerin2010novel} observed in PET-based simulation studies that incorporating the energy attribute while performing scatter correction reduced the bias in the activity distribution estimates by up to 40\% for a single-scatter simulation model. 
However, conventional QSPECT methods are typically designed to reconstruct the activity uptake distributions from binned projections, and thus, the LM data, including the energy attribute, is binned to enable processing with these methods. 
A method that processes the data in LM format, including the energy attribute, may provide a mechanism to improve performance of QSPECT for $\alpha$-RPTs.

Based upon these premises and motivated by the goal of addressing the challenges of $\alpha$-RPT SPECT, in this manuscript, we propose a low-count-SPECT reconstruction method for multi-emission-peak isotopes that uses data from multiple energy windows (MEW) in LM format and including the energy attribute of each detected photon. The proposed method, referred to as LM-MEW SPECT reconstruction method, compensates for scatter, attenuation, and collimator-detector response. The method is based on inverting the physics of the forward model of the LM data acquisition process. Thus, as part of designing the method, we first develop a model for the acquisition of LM data specifically in the context of multi-emission-peak isotopes. The model advances upon a previously proposed approach \cite{jha2013retrieving,rahman2020fisher}, but while the previous approach considered only isotopes with a single photopeak, the advanced approach proposed here accounts for isotopes having multi-peak emission spectra. Following the design of the reconstruction method, we develop an ordered-subsets version of the method and implement it using a parallelized multi graphics processing unit (GPU) routine for computational efficiency. The method is then objectively evaluated on the quantitative task of estimating activity uptake within known regions in the context of $\alpha$-RPT SPECT with [$^{223}$Ra]RaCl${_2}$.

\section{Method: Theory and implementation}
\label{sec:path_EM}
The central idea of the proposed method is to estimate the activity distribution that maximizes the likelihood of the LM data detected by the SPECT system across multiple energy windows, where the LM data includes the energy attribute. We begin by deriving an expression for the likelihood of this data. To maximize the high-dimensional likelihood, we derive an expectation-maximization algorithm. For computational efficiency, we then formulate an ordered-subsets version of this technique.
\subsection{Problem formulation}
Consider a SPECT system that is imaging a patient administered with radiotracer containing multi-emission-peak isotope and acquires data in LM format. While the isotope distribution is continuous, for this reconstruction problem, we represent the isotope distribution in a voxelized grid space, denoted by a $Q$-dimensional vector $\f$. Let $\lambda_q$ denotes the activity at the $q^\text{th}$  voxel. 
The isotope emits photons at multiple emission energies. Denote the probability \rchange{of emitting photons at energy $E_0^w$}{for a decay to result in an emitted photon having a specific energy $E_0^w$} by $\alpha_w$ and the total number of emission energies by $W$.

Next, consider that the SPECT system acquires data for a fixed scan time (preset-time system) (the proposed analysis can easily be extended to a preset-count system).
In the measurement time $T$, denote the number of LM events detected by $J$. 
For each event, the position of interaction of the photon with the crystal, the energy deposited by the event at the interaction site, and the detector angle, are recorded.
Denote the true and estimated attributes of an event $j$ as the vectors $\bA_j$ and $\bAhat_j$, respectively. Thus, the observed data measurements consist of the set of measured attribute vectors $\hat{\mathcal{A}} = \{ \bAhat_j, j = 1, 2, \ldots J \}$ and the number of detected events $J$, collected over a broad energy spectrum. 
Given this setting, the reconstruction problem is to estimate $\f$ given the measured data $\hat{\mathcal{A}}$. 

\subsection{Likelihood of list-mode data using path-based formalism}
To derive the likelihood of the measured data, we note that the $J$ detected events are all independent. Thus, the likelihood of the measured LM data is given by
\begin{align}
\pr(\hat{\A},J|\obj) &= \Pr(J|\obj)\pr(\hat{\A}|\obj)\nonumber\\
 &= \Pr(J|\obj)\prod_{j=1}^{J}\pr(\bAhat_j|\obj),
\label{eq:lm_ll_gen}
\end{align} 
where $\Pr(.)$ and $\pr(.)$ denote the probabilities of discrete and continuous random variables, respectively. 
For a preset time $T$, $J$ is Poisson distributed. We denote the mean rate of detected photons in the SPECT system as $\beta$, so the probability of the detected number of events is given by
\begin{align}
\Pr(J|\obj)= \dfrac{(\beta T)^J}{J!}\exp(-\beta T).
\label{eq:j_poiss}
\end{align}
Next, we need to obtain an expression for the term $\pr(\bAhat_j | \obj)$, i.e. the probability of detecting the $j^{\th}$ LM event given $\f$. 
Due to the complexity in physically modeling the detection of a LM event, obtaining an analytical expression for this term is challenging. 
To address this issue, note that each LM event is the result of a photon being emitted from a certain location, traveling in a certain direction, in some cases, scattering at certain locations, and finally being incident on the detector.
In other words, any LM event is the result of a photon traversing a particular path before being detected (Fig.~\ref{fig:spect_system}).
Thus, we can decompose the probability of an event as a mixture model over all possible paths \cite{jha2013retrieving,rahman2020fisher}.
To formalize this mathematically, let $\pr(\bAhat_j|\P,E_0^w)$ denote the probability that the $j^\text{th}$ event is detected given that the event was the result of a photon being emitted at an energy $E_0^w$ and traversed a particular path $\P$. 
Let $\Pr(\P,E_0^w|\obj)$ denote the probability that a photon is emitted at energy $E_0^w$ and follows a path $\P$.
Based on the path-based decomposition, we can write the probability of a LM event as the following mixture model: 
\begin{align}
\pr(\bAhat_j|\obj)=\sum_{w=1}^{W}\sum_{\P}\pr(\bAhat_j|\P,\E_0^w)\Pr(\P,\E_0^w|\obj).
\label{eq:Aj_expansion}
\end{align}
The expression for the term $\pr(\bAhat_j|\P,\E_0^w)$ was calculated based on the energy and position resolution of the detector \cite{jha2013retrieving,rahman2020fisher}. In the following sections, we provide a brief description for deriving the expression for the term $\Pr(\P,\E_0^w|\obj)$ and the log-likelihood of the observed LM data. The term $\Pr(\P,\E_0^w|\obj)$ expresses the probability of detecting a photon transferred through a path and thus reflects the physical model used by the path-based approach.
\subsubsection{Expression for the probability of detecting a photon transferred through a path}
The term $\Pr(\P,\E_0^w| \obj)$ denotes the probability to detect a photon that is transferred through the path $\P$ when the emission energy is $\E_0^w$. 
More specifically, this term is the ratio of mean rate of photons incident on detector through the considered path $\P$ with emission energy $E_0^w$ to mean rate of photons incident on detector \cite{jha2013retrieving,rahman2020fisher,jha2013joint,barrett1997list}.
Following a similar approach as in \cite{jha2013retrieving,rahman2020fisher,jha2013joint,barrett1997list}, we can derive that
\begin{align}
\Pr(\P,\E_0^w|\obj)=\frac{\lambda_{\mathcal{E}}({\P},\E_0^w)s(\P,\E_0^w)}{\sum_{w=1}^{W}\sum_{\P'}\lambda_{\mathcal{E}}({\P'},\E_0^w)s(\P',\E_0^w)},
\label{eq:pr_path_1}
\end{align}
where $\lambda_{\mathcal{E}}(\P,\E_0^w)$ denotes the rate of photon emission \rnew{at energy $E_0^w$ }in the voxel from which the path $\P$ originates\rold{ at energy $E_0^w$}, and $s(\path,\E_0^w)$ denotes the sensitivity of the detector for photons emitted at energy $E_0^w$ and traversing the path $\P$. 
Consider that the path $\P$ describes the trajectory of photons that are emitted from location $\r_0$ with energy $E_0^w$, scatter $n$ times at locations $\r_1, \ldots, \r_n$,  and get detected at the location $\r_d$ (Fig.~\ref{fig:spect_system}).
Denote the energy of the photon after each of these scattering events by $E_1^w, \ldots, E_n^w$, respectively.
Then the expression for $s(\P,E_0^w)$ is given by \cite{jha2013retrieving,rahman2020fisher}
\begin{align}
s(\P,E_0^w)&=\frac{\Delta \Omega}{4\pi}\exp\left\{-\gamma(\r_0, \r_1, E_0^w) - \ldots -\gamma(\r_n, \r_d, E_n^w)\right\}\nonumber \\
&\times t(\r_n, \tht_{k_n}) \prod_{m=1}^{n}\Delta_{q_m}(\subpath_{\r_{m-1}, k_{m-1}}) \nonumber \\
&\times \prod_{m=1}^{n}K(\tht_{k_{m-1}}, \tht_{k_{m}}, E_{m-1}^w| \r_m),\label{eq:s_p}
\end{align}
where, the term $\gamma(\r_u, \r_v, E_i)$ denotes the path integral between locations $\r_u$ and $\r_v$ at the energy $E_i$, and is given by
\begin{align}
\gamma(\r_u, \r_v, E_i) = \int_{0}^{|\r_u-\r_v|}\mu\left(\r_u-t\frac{\r_u-\r_v}{|\r_u-\r_v|}, E_i\right)~dt,
\label{eq:radpath}
\end{align}
where $\mu(\r, E)$ denotes the attenuation coefficient at the location $\r$ and energy $E$. 
Further, the term $\subpath_{\r_u, k_v}$ in Eq.~\eqref{eq:s_p} denotes a discretized subpath within the path $\P$ and describes the unit of space where radiation propagates between one scatter voxel to another. 
Each subpath $\subpath_{\r_u, k_v}$ is a right-angular cone with a solid angle $\Delta\Omega$ that has its apex located at the location $\r_u$ and its axis at an angular direction $\tht_{k_v}$ \cite{rahman2020fisher}. 
The term $t(\r_n, \tht_{k_n})$ denotes the sensitivity of the collimator for a photon emitted or scattered from the location $\r_n$ and traversed along the direction $\tht_{k_n}$. 
The term $\Delta_{q_m}(\subpath_{\r_u, k_v})$ denotes the intersection length of a voxel $q_m$ with axis of the subpath $\subpath_{\r_u, k_v}$, where the voxel $q_m$ contains the scattering location $\r_m$. 
The term $K(\tht_u, \tht_v, E_i | \r_j)$ denotes the differential scattering cross section at the location $\r_j$ with incoming direction $\tht_u$ and outgoing direction $\tht_v$ for a photon with energy $E_i$ before scattering. This term can be calculated using the Klein-Nishina formula \cite{klein1929streuung}.
For simplicity, we assume that the sub-paths always start from center of the voxels. 
We also assume that the attenuation coefficient and the energy have piecewise-linear relationship.

Next, note that the denominator in Eq.~\eqref{eq:pr_path_1} represents the mean rate of detected photons, so that 
\begin{align}
\beta = \sum_{w=1}^{W}\sum_{\P}\lambda_{\mathcal{E}}({\P},E_0^w)s(\P,E_0^w).
\label{eq:th_beta}
\end{align}
\subsubsection{Expression for the log-likelihood}
Using Eqs.~\eqref{eq:lm_ll_gen}-\eqref{eq:pr_path_1} and \eqref{eq:th_beta}, we can write the log-likelihood of observed LM data as
\begin{align}
\L(\obj|\hat{\A},J)=\sum_{j=1}^{J}\log\left(\sum_{w=1}^{W}\sum_{\P}\pr(\bAhat_j|\P,E_0^w)\lambda_{\mathcal{E}}({\P},E_0^w)s(\P,E_0^w)\right) +\nonumber\\
J\log( T)-T\sum_{w=1}^{W}\sum_{\P}\lambda_{\mathcal{E}}({\P},E_0^w)s(\P,E_0^w) - \log J!.
\label{eq:lll_med}
\end{align}

\subsection{Expectation-maximization (EM) algorithm}
To estimate $\obj$ using a maximum-likelihood approach, we need to differentiate the log-likelihood expression in Eq.~\eqref{eq:lll_med}, which is challenging. Here, we take advantage of the fact that every detected LM event follows a certain path. While we do not know the path that the photon has taken, this fact lends this problem to an expectation-maximization (EM)-based solution. 
Advancing on the treatment in Shepp \textit{et al} \cite{shepp1982maximum}, Lange \textit{et al} \cite{lange1984reconstruction}, Parra and Barrett \cite{parra1998list} and Khurd \textit{et al} \cite{khurd2004globally}, we first define a hidden variable $z_{j,\P,w}$ for each event and each path, where 
\begin{align}
 z_{j,\P,w} = 
  \begin{cases} 
   1, & \text{ if event $j$ had an emission energy of $E_0^w$ and took the path $\P$}. \\
   0,       & \text{ otherwise}.
  \end{cases}
\label{eq:def_hidden_vector}
\end{align}
Thus, for each event $j$, we can define a hidden vector $\z_j$ where each element of this vector indicates a unique path $\P$ and a unique emission energy $E_0^w$. Because an event can not traverse more than one path, only one of the elements of the vector $\z_j$ will be 1. The observed LM data, in conjunction with this hidden vector $\z_j$, form the complete data. The likelihood of this complete data is given by
\begin{align}
&L_C(\obj | \{\bAhat_j,\z_j;j=1,\ldots J\},J)\nonumber\\ &= \pr(\{\bAhat_j,\z_j;j=1,\ldots J\},J|\obj)\nonumber\\
&=\Pr(J|\obj)\prod_{j=1}^{J}\pr(\bAhat_j,\z_j|\obj)\nonumber\\
&=\Pr(J|\obj)\prod_{j=1}^{J}\pr(\bAhat_j|\z_j,\obj)\Pr(\z_j|\obj),
\label{eq:likelihood_complete}
\end{align}
where in the third step, we have used the fact that the $J$ LM events are independent, and in the fourth step, we use the chain rule of probability. 
Next, using the path-based decomposition (Eq.~\eqref{eq:Aj_expansion}) and the properties of $\z_j$ (Eq.~\eqref{eq:def_hidden_vector}), we obtain  
\begin{align}
\pr(\bAhat_j|\z_j,\obj) &= \sum_{w=1}^{W}\sum_\P \pr(\bAhat_j|\P,E_0^w)\Pr(\P,E_0^w|\z_j,\obj) \nonumber\\
&= \sum_{w=1}^{W}\sum_\P z_{j,\P,w}\pr(\bAhat_j|\P,E_0^w) = \prod_{w=1}^{W}\prod_\P \pr(\bAhat_j|\P,E_0^w)^{z_{j,\P,w}}.
\label{eq:pr_Ajhat_zj}
\end{align}
In the second step, we have used the definition of $z_{j,\P,w}$. In transitioning from the second to the third step, we have used the fact that $z_{j,\P,w}$ is 1 for a specific path and emission energy, and 0 otherwise. Thus, the summation can be replaced by a multiplication. Similarly,
\begin{align}
\Pr(\z_j|\obj) &= \sum_{w=1}^{W}\sum_\P \Pr(\z_j|\P, E_0^w) \Pr(\P,E_0^w|\obj)\nonumber\\
&= \sum_{w=1}^{W}\sum_\P z_{j,\P,w} \Pr(\P,E_0^w|\obj) = \prod_{w=1}^{W}\prod_\P \Pr(\P,E_0^w|\obj)^{z_{j,\P,w}},
\label{eq:pr_zj_lambda}
\end{align}
following the same rationale as used while deriving Eq.~\eqref{eq:pr_Ajhat_zj}.
Inserting the expressions derived in Eqs.~\eqref{eq:pr_Ajhat_zj} and \eqref{eq:pr_zj_lambda} into Eq.~\eqref{eq:likelihood_complete}, taking the logarithm of the likelihood, inserting the expressions from Eqs.~\eqref{eq:j_poiss} and \eqref{eq:pr_path_1} for $\Pr(J|\obj)$ and $\Pr(\P,\E_0^w|\obj)$ respectively, and then finally inserting the expression for $\beta$ from Eq.~\eqref{eq:th_beta}, we derive the complete data log-likelihood as
\begin{align}
&\L_C(\obj|\{\bAhat_j,\z_j;j=1,\ldots J\},J)\nonumber\\
&=\sum_{j=1}^{J}\left[\sum_{w=1}^{W}\sum_{\P}z_{j,\P,w}\left\{\log \pr(\bAhat_j|\P,E_0^w)+\log \lambda_{\mathcal{E}}({\P},E_0^w) + \log s(\P,E_0^w)\right\}\right] \nonumber\\
& \quad\quad + J \log T-T\sum_{w=1}^{W}\sum_{\P}\lambda_{\mathcal{E}}({\P},E_0^w)s(\P,E_0^w) - \log J!.
\label{eq:cd_lll}
\end{align}
In the expectation (E) step, we take the expectation of the log-likelihood conditioned on observed data. 
Since the expression is linearly related to $z_{j,\P,w}$, this is equivalent to replacing $z_{j,\P,w}$ in Eq.~\eqref{eq:cd_lll} with its expected value conditioned on the observed data, denoted by $\bar{z}_{j,\P,w}$.
Then, the conditional expectation of the complete data log-likelihood  is given by
\begin{align}
&\bar{\L}_C  =\sum_{j=1}^{J}\left[\sum_{w=1}^{W}\sum_{\P}\bar{z}_{j,\P,w}\left\{\log \pr(\bAhat_j|\P,E_0^w)+\log \lambda_{\mathcal{E}}({\P},E_0^w) + \log s(\P,E_0^w)\right\}\right] \nonumber\\
& \quad\quad + J \log T-T\sum_{w=1}^{W}\sum_{\P}\lambda_{\mathcal{E}}({\P},E_0^w)s(\P,E_0^w) - \log J!.
\label{eq:cd_lll_ex}
\end{align}
Let $\lambda({\P})$ denote the activity in the voxel from which the path $\P$ originates. Thus, $\lambda_{\mathcal{E}}({\P},E_0^w) = \alpha_w\lambda({\P})$.
In the maximization (M) step, we take the derivative of the log-likelihood (Eq.~\eqref{eq:cd_lll_ex}) with respect to $\lambda_q$. Setting that to zero yields
\begin{align}
0&=\dfrac{1}{\lambda_q}\sum_{j=1}^{J}\sum_{w=1}^{W}\sum_{\P_q}\bar{z}_{j,\P_q,w} - T\sum_{w=1}^{W}\alpha_w\sum_{\P_q}s(\P_q,E_0^w),
\label{eq:lm_mlem_d}
\end{align}
where $\P_q$ denotes the set of paths that start from voxel $q$.
This yields the following iterative update equation at the  $(i+1)^{\th}$ iteration:
\begin{align}
\lambda_q^{(i+1)}=\dfrac{\sum_{j=1}^{J}\sum_{w=1}^{W}\sum_{\P_q}\bar{z}_{j,\P_q,w}^{(i+1)}}{T\sum_{w=1}^{W}\alpha_w\sum_{\P_q}s(\P_q,E_0^w)},
\label{eq:lm_mlem}
\end{align}
where 
\begin{align}
\bar{z}_{j,\P,w}^{(i+1)}&= \Pr\left( \P,E_0^w |\bAhat_j, \obj^{(i)}\right)\nonumber\\
&=\dfrac{\pr(\bAhat_j|\P,E_0^w)\Pr(\P,E_0^w|\f^{(i)})}{\sum_{w=1}^{W}\sum_{\P'}\pr(\bAhat_j|\P',E_0^w)\Pr(\P',E_0^w|\f^{(i)})}.\label{eq:z_bar_update}
\end{align}
In deriving $\bar{z}_{j,\P,w}^{(i+1)}$, we have used the fact that $z_{j,\P,w}$ can take values of either 0 or 1.  
Intuitively, the numerator in Eq.~\eqref{eq:lm_mlem} represents the expected number of detected events originating from voxel $q$. The denominator represents the sensitivity of the system at voxel $q$ multiplied with the scan time period $T$. Thus, the right side of the equation represents the activity at voxel $q$. 

\subsection{List-mode ordered-subsets expectation-maximization algorithm (LM-OSEM)}\label{sec:os-lm-mlem}
The iterative process described by Eq.~\eqref{eq:lm_mlem} is computationally intensive, especially as the numerator in Eq.~\eqref{eq:lm_mlem} is challenging to compute. To improve the computational efficiency, we develop an ordered-subsets (OS) approach advancing on the treatment in Hudson \textit{et al} \cite{hudson1994accelerated} and Khurd \textit{et al} \cite{khurd2004globally}. 

The ordered subsets are formed based on the angle of detection of the LM events. Denote $S_s^j$ as the subset of events that are acquired in detection angles defined by the $s^\text{th}$ subset. Denote $S_s^\P$ as the subset of paths that can result in events that can be detected in angles defined by the $s^\text{th}$ subset. 
Denote the number of subsets and iterations by $N_s$ and $N_g$, respectively.
Then the update equation for the $s^\text{th}$ subset and the $(i+1)^\text{th}$ iteration is given by
\begin{align}
\lambda_q^{(i+1,s)}=\dfrac{\sum\limits_{\substack{j=1\\j\in S^j_s}}^{J}\sum_{w=1}^{W}\sum\limits_{\P_q\in S_s^{\P}}\bar{z}_{j,\P_q,w}^{(i+1,s)}}{T\sum_{w=1}^{W}\alpha_w\sum\limits_{\P_q\in S_s^{\P}}s(\P_q,E_0^w)}&;i=1,\ldots,N_g\nonumber\\
&,s=1,\ldots,N_s,
\label{eq:os-lm-mlem}
\end{align}
where
\begin{align}
\bar{z}_{j,\P_q,w}^{(i+1,s)} &=\dfrac{\pr(\bAhat_j|\P_q,E_0^w)\Pr(\P_q,E_0^w|\f^{(\boldsymbol{\eta})})}{\sum_{w=1}^{W}\sum_{\P' \in S_s^\P}\pr(\bAhat_j|\P',E_0^w)\Pr(\P',E_0^w|\f^{(\boldsymbol{\eta})})}; j\in S_s^j \nonumber\\
&= \dfrac{\pr(\bAhat_j|\P_q,E_0^w)\lambda^{(\pmb{\eta})}_{\mathcal{E}} (\P_q,E_0^w) s(\P_q,E_0^w)}{\sum_{w=1}^{W}\sum_{\P' \in S_s^\P}\pr(\bAhat_j|\P',E_0^w) \lambda^{(\pmb{\eta})}_{\mathcal{E}} (\P',E_0^w) s(\P',E_0^w)}; j\in S_s^j,
\label{eq:z_bar_update_os}
\end{align}
where in the second step, we used Eq.~\eqref{eq:pr_path_1} and where $\boldsymbol{\eta}$ is a $2$-D vector, the first element of which denotes the iteration number and the second element denotes the subset index. Let $\f^{0}$ denote the initial activity map input to the proposed iterative procedure. Then
\begin{align}
\boldsymbol{\lambda^\eta} = \begin{cases}
\f^{0} &,\text{if } i=1 \text{ and } s=1.\\
\f^{(i,N_s)} &,\text{if } i \neq 1 \text{ and } s = 1.\\
\f^{(i+1,s-1)} &,\text{otherwise.}
\end{cases}
\end{align}
In Eq.~\eqref{eq:z_bar_update_os}, we have used the fact that $\pr(\bAhat_j|\P,E_0^w)=0$ when $j \in S_s^j$ and $\P\notin S_s^\P$ while expressing the denominator.

An asymptotic analysis of the computational requirements of the LM-OSEM algorithm is given in Appendix 8.

\subsection{GPU-based implementation of the LM-OSEM Algorithm}
To further improve the computational efficiency of the algorithm, we implemented it on multiple graphics processing units (GPUs). Pseudo-code for the algorithm is given in Algorithm 1 within Appendix 7, with a more detailed description in Appendix 9. The proposed MEW-LM-OSEM reconstruction method was implemented on a system with an Intel Xeon processor and four NVIDIA TESLA V100 GPUs, each with 32GB of RAM. Note that the implemented algorithm can perform reconstruction on any arbitrary number of GPUs available on the system.

\section{Evaluation of the proposed method}
We evaluated the proposed method on the task of estimating activity uptake within a defined region in the context of a simplified $\ra_223$-based $\alpha$-RPT SPECT setup. This evaluation requires an experimental setup where the ground truth uptake is known. Given this need, we designed this to be a simulation study where we imaged a synthetic phantom containing uptake of the $\ra_223$. Since the method is highly computationally intensive even after GPU acceleration, we restrict these studies to 2D, and model only one degree of scatter. The overall goal of the evaluation study was to assess whether the proposed method can yield improved quantification performance in this 2D setting. If so, this would motivate implementation of this method for more realistic 3D settings. We first describe the components of our evaluation study.

\subsection{Overall evaluation framework}
\label{sec:overall_eval}
\subsubsection{Synthetic phantom model}
\label{sec:synth_phan}
The synthetic activity phantom (Fig.~\ref{fig:synth_true_act_attn}a) was elliptical with major and minor radii of 11 cm and 9 cm, respectively. The phantom had four circular regions, with radii 7 mm, 10 mm, 12 mm and 14 mm. Each region was centered symmetrically with respect to the center of rotation of the imaging system. These circular regions served as the signals, simulating tumors in a patient. The activity inside the circular regions was fixed. We chose the small circular regions with radii 7 mm and 10 mm to evaluate the quantification performance for regions that are close to the SPECT system resolution. To simulate patient variability, background activity was modeled as a stochastic clustered lumpy model \cite{bochud1999statistical}. The signal-to-background ratios (SBRs) in the circular regions were assigned to be 2:1, 4:1, and 6:1 in different experiments.
The attenuation map of the phantom was modeled as in Fig.~\ref{fig:synth_true_act_attn}b, consisting of a rim and an inner region with attenuation coefficients of $0.21~\cmtr^{-1}$ and $0.18~\cmtr^{-1}$, respectively, at 85 keV. \rchange{The two attenuation coefficients model two different types of soft tissue. }{In choosing these two attenuation coefficients, the material composition was set to water and the density varied. }To model the energy dependence of the attenuation coefficient, we assumed a piecewise-linear relationship between attenuation coefficient and energy. More specifically, we assumed a bilinear curve to approximate the energy dependence. The slope of the curve was determined using the attenuation coefficient of water at different energy values. This bilinear assumption is approximately satisfied over the wide energy spectrum of radium-223 isotope considered in this evaluation study.

\subsubsection{SPECT simulation}
\label{sec:spect_sim}
We simulated a 2-D SPECT system with a geometry similar to the GE Optima NM/CT 640 SPECT/CT system. Medium-energy general-purpose parallel-hole (MEGP) collimator was selected for this study with specifications similar to the MEGP collimator used in the above-mentioned GE scanner. The isotope was considered to be $\ra_223$, which emitted photons  at the energy values 81.1 keV, 83.8 keV, 95 keV, 144 keV, 155 keV and 270 keV with probabilities according to \cite{hindorf2012quantitative}. A NaI scintillation detector was simulated, with an intrinsic position resolution of 4 mm. The overall system resolution was 9.8 mm at 10 cm depth. The energy resolution of the detector was set to 10\% FWHM at 140 keV and was assumed to have an energy dependence proportional to the inverse of the square root of the energy value \cite{prince2006medical}. We used a previously validated MC code \cite{rahman2020fisher} to generate LM events that had scattered at most once. The LM data was acquired in 120 evenly spaced fixed angular positions over $360^{\circ}$. A circular field-of-view with a 30 cm diameter was considered. The LM data for each detected event contained the measured position of the interaction in the detector, the energy deposited in the detector element, and the angle of acquisition. We set the acquisition time to generate approximately $5 \times 10^3$ events in the energy window spanning from 68 keV to 102 keV. This count constitutes approximately 50\% of the total counts acquired in all three energy windows in the ranges of 68-102 keV, 123-184 keV and 243-297 keV.

\subsubsection{Procedure to perform quantification task}
The method was applied to the LM data generated using our simulated SPECT system. The reconstruction method was executed with four subsets and 16 iterations, because it was observed that for this configuration, the estimated activity in the circular regions almost converged. The reconstruction was performed over a $64\times 64$ grid, with a pixel size of 4.6 mm. In the reconstruction, first-order scatter was modeled. From the reconstructed activity image, we estimated the activities in the different circular regions by taking the sum of the activity in the different voxels within each region of interest.

These experiments were conducted over multiple realizations. For the synthetic phantom, in each realization, a different statistical sample of the object background was considered, thus simulating patient variability. Further, for each object realization, the LM data was generated multiple times independently, thus modeling variability introduced by system noise. 

\subsubsection{Figure of merit for evaluation}
The reliability (accuracy and precision) of the estimated uptake in the circular region was quantified using the ensemble normalized root mean square error (ENRMSE). This metric was chosen as it quantifies both accuracy and precision. Let $S$ denote the number of object realizations, $N$ denotes the number of noise realizations, $y_{sk}$ denotes the true activity uptake inside the $k^{\th}$ circular region for $s^{\th}$ object realization, and $\hat{y}_{sk}^n$ denotes the estimated activity uptake in the $k^{\th}$ circular region of the $n^{\th}$ noise realization and $s^{\th}$ object realization. The ENRMSE of the estimated uptake in $k^{\th}$ circular region was computed as
\begin{align}
\text{ENRMSE}_{k} = \dfrac{1}{S}\sum_{s=1}^{S}\left\{\dfrac{1}{y_{sk}}\sqrt{\dfrac{\sum_{n=1}^{N}(\hat{y}_{sk}^{n}-y_{sk})^2}{N}}\right\}.
\end{align}
We also computed the ensemble normalized bias for quantifying the accuracy of the proposed method on the task of estimating circular-region uptake. The ensemble normalized bias of the estimated uptake in $k^{\th}$ circular region was computed as

\begin{align}
\text{Ens. Norm. Bias}_{k} = \dfrac{1}{SN}\sum_{s=1}^{S}\left\{\dfrac{\sum_{n=1}^N(\hat{y}_{sk}^{n}-y_{sk})}{y_{sk}}\right\}.
\end{align}

To quantify precision, we computed the ensemble normalized standard deviation, which was computed for the $k^{\th}$ region as follows:
\begin{align}
\text{Ens. Norm. Std.}_{k} = \dfrac{1}{S}\sum_{s=1}^{S}\left\{\dfrac{1}{y_{sk}}\sqrt{\dfrac{\sum_{n=1}^{N}|\hat{y}_{sk}^{n}-\frac{1}{N}\sum_{n'=1}^{N}\hat{y}_{sk}^{n'}|^2}{N-1}}\right\}.
\end{align}

\subsection{Experiments}
\subsubsection{Agreement between path-based modeling approach and MC simulation}\label{sec:agreement}
To evaluate the accuracy of the path-based modeling approach (Sec.~\ref{sec:path_EM}) in the context of modeling LM acquisition of 2-D SPECT of $\alpha$-RPTs within a multi-energy window setup in a single-scatter setting, we assessed the agreement between this modeling approach with a corresponding MC-based simulation. To perform the MC-based simulation, we used an in-house MC software \cite{rahman2020fisher} modified for $\alpha$-RPT SPECT.

For this evaluation study, we used the activity and attenuation phantom described in Sec.~\ref{sec:synth_phan}. In the activity phantom, the SBR value of each of the circular regions was set at 4:1.  Using this MC software, we determined the number of detected events that originated from a specific pixel, had a fixed emission energy, and had scattered at most once. In the MC process, single-order scatter was modeled using Klein-Nishina formula normalized to a 2-D plane. The MC software simulated the SPECT system described in Sec.~\ref{sec:spect_sim}. We generated $\sim 8\times 10^4$ counts through this MC process. This number of counts is ~10x counts compared to the counts acquired in a typical $\alpha$-RPT SPECT protocol because we needed to approximate the sensitivity of each pixel. We describe this evaluation study in more detail in the next paragraph. Note that, in later experiments, we acquired counts typically seen in $\alpha$-RPT protocol to simulate true low-count setting. The MC-generated data was collected in three energy windows in the ranges of 68-102 keV, 123-184 keV and 243-297 keV. More details of the energy window configurations are discussed later in this section. 

Using the MC software-based measurements, we generated a sensitivity map where each pixel in the sensitivity map denoted the number of detected LM events (unscattered and first-order scatter) that originated from that pixel and with the considered emission energy. Thus, six sensitivity maps were generated corresponding to each of the six emission energies as mentioned in Sec.~\ref{sec:spect_sim}. Then, we used the proposed path-based modeling to generate the corresponding sensitivity map for each emission energy. We compared the sensitivity map generated from MC simulation and path-based modeling to evaluate the agreement between these approaches.

\subsubsection{Evaluating performance for different circular-region sizes}
Using the simulation setup as described in Sec.~\ref{sec:overall_eval}, we evaluated the performance of the proposed approach on the task of estimating activity uptake within different-sized regions. For these experiments, we set the SBR of each of the circular regions in the synthetic phantom at 4:1. The proposed method uses data from multiple energy windows. Specifically, we considered three non-overlapping energy windows in the ranges of 68-102 keV, 123-184 keV and 243-297 keV. We denote these energy windows by EW1(PP), EW2(PP) and EW3(PP), respectively. The PP stands for photopeak and is named such since the energy window encompasses an emission peak. We denote the multiple energy window (MEW) configuration of our proposed method by EW1+EW2+EW3(PP). 

We compared the proposed method to a method where only a single energy window was used (LM-SEW). This method used only the measurements acquired in EW1(PP).  Conducting this experiment evaluated the efficacy of using multiple energy windows. We also compared the proposed approach to a more conventional reconstruction method where energy attribute was binned and only data from EW1(PP) was considered (Binned-SEW), similar to multiple other previous studies \cite{gustafsson2020feasibility,benabdallah2019223ra,li2021projection,owaki2017ra}. This comparison study, in addition to evaluating performance against a more traditional reconstruction approach, also evaluated the efficacy of using energy attribute in LM format. To implement this method, all LM events within EW1(PP) were assigned the fixed energy value of 85 keV. The position attribute was not binned to ensure that we could specifically study the impact of binning the energy attribute. This data was then reconstructed using the proposed reconstruction approach. 

Both the LM-SEW and Binned-SEW approaches were OSEM-based methods that were implemented as special cases of the proposed method and compensated for attenuation, first-order scatter and collimator-detector response. The similarity in implementation was to ensure that the comparison study was rigorous and specifically studied the impacts of using multiple energy windows and binning of the energy window.

\subsubsection{Evaluating performance as a function of contrast of the circular region}
In these studies, we varied the activity in the circular regions in the synthetic phantom. The mean value of the SBR was assigned to one of three ratios: 2:1, 4:1, or 6:1. Following the experimental setup as described in Sec.~\ref{sec:overall_eval}, we evaluated the performance of the proposed method on the task of estimating activity uptake in the circular regions. For this experiment, we used the MEW configuration that uses all three energy windows. 

\section{Results}
\subsection{Agreement between path-based modeling approach and MC simulation}
Fig.~\ref{fig:mc_valid} shows the sensitivity maps obtained with the MC simulation and the proposed path-based modeling approach for the six different emission energies. The normalized difference between these two methods is also shown. The mean normalized difference between the sensitivities obtained with the methods was close to 2.5\% over all emission energies. This provides evidence that the proposed path-based formalism was yielding an accurate modeling of the considered 2-D SPECT system imaging an isotope with multiple emission peaks and with at most single-order scatter. 

\subsection{Evaluating performance for different circular-region sizes} 
Fig.~\ref{fig:comp_all} shows the ensemble normalized bias, ensemble normalized standard deviation and ENRMSE as a function of circular-region radius for the proposed LM-MEW method, LM-SEW method, and Binned-SEW method. We observe that the ENRMSE and ensemble normalized bias between true and estimated uptake are lower for the proposed LM-MEW method compared to the more traditional LM-SEW and Binned-SEW methods. Moreover, we observe that the ensemble normalized standard deviation value is lower for the proposed LM-MEW method compared to the LM-SEW and Binned-SEW methods. We also observe, in general, that as the circular-region radius increases, ensemble normalized bias, ensemble normalized standard deviation and ENRMSE decrease for all the methods.

\subsection{Evaluating performance as a function of contrast of the circular region} 
The ensemble normalized bias, ensemble normalized standard deviation and ENRMSE as a function of SBR and circular-region radius for the proposed method are shown in Fig.~\ref{fig:diftum}. In general, as the SBR increases, the ENRMSE and ensemble normalized standard deviation values decrease. Similar to all the previous results, ensemble normalized bias, ensemble normalized standard deviation and ENRMSE decrease in general as the circular-region radius increases.

\section{Discussion}
In this manuscript, being motivated by the problem of performing quantitative SPECT for $\alpha$-RPTs, we proposed a low-count LM-MEW SPECT reconstruction method where the isotopes have multiple emission peaks. The method uses data in LM format and multiple energy windows, including the energy attribute. The method models relevant image-degrading processes in SPECT, including attenuation, scatter, and collimator-detector response. We evaluated the method on the task of estimating activity uptake within known circular regions of a 2-D phantom for a $\ra_223$-based $\alpha$-RPT setup being imaged by 2-D SPECT system within a single-scatter simulation setting. The results in Fig.~\ref{fig:comp_all} provide evidence that the proposed method outperforms strategies that only use single energy window or bin the energy attribute into a single window. Our results suggest that the inclusion of photons in LM format and using photons from multiple energy windows may result in an improvement in quantification performance.

In the results of Fig.~\ref{fig:comp_all}, we observe that including data from multiple energy windows yielded improved estimation of activity uptake in circular regions of all considered sizes. This indicates the use of photons in multiple energy windows is beneficial for quantifying the activity uptake in regions with different sizes. By including more energy windows, we used more of the detected photons, leading to an increased effective system sensitivity. Our result shows that this increased effective sensitivity translates to an improvement in the quantification task. This increase in sensitivity is important in the context of SPECT with $\alpha$-RPTs, considering the low count levels in those imaging applications. Further, we observe that the bias in the estimated activity decreased significantly (Fig.~\ref{fig:comp_all}) when multiple energy windows were used.  We hypothesize that the reduction in bias is because the inverse problem is less ill-posed when using increased number of measurements from multiple energy windows.

Results in Fig.~\ref{fig:comp_all}a demonstrated that for the SEW configuration, processing the energy attribute in LM format led to a reduction in the ensemble normalized bias of estimated uptake as compared to binned format. It is known that binning attributes leads to loss of information \cite{jha2015singular}. Our results show that the added information gained by processing data in LM format also translates to an improvement in performance on quantification tasks.  Previous studies have shown that binning the position and angular attribute leads to an increase in bias for the task of absolute quantification in a region of interest \cite{jha2015estimating}. In this study, we continue to observe a similar finding even for the binning of energy attribute.
However, we do observe in Fig.~\ref{fig:comp_all}b that, for the SEW configuration, processing data in LM format leads to higher ensemble normalized standard deviation compared to binned format. A similar observation was found in the study conducted by Jha \textit{et al} \cite{jha2015estimating} where the angle of detection was binned. The lower precision could be attributed to the fact that the experiments are conducted in a low-count setting \cite{jha2015estimating}. However, despite that, when taking both accuracy and precision into account (Fig.~\ref{fig:comp_all}c), we observe that LM format leads to lower ENRMSE value compared to binned format.

The performance of the proposed method on the quantification task was observed to depend on the size of the circular region and the SBR value. Fig.~\ref{fig:diftum} shows that an increase in the circular-region size resulted in a decrease in the ENRMSE value of the estimated uptake. This is an expected finding since as the circular-region size increases, the impact of the limited spatial resolution and resultant partial volume effects (PVEs) on quantification performance is known to decrease. Another expected finding was that, in general, an increase in the SBR value caused the ENRMSE value to decrease. 
We also observe from Fig.~\ref{fig:diftum}a that a decrease in SBR values resulted in a decrease in bias. This is consistent with observations in a study by Cloquet \textit{et al} \cite{cloquet2011does} that looked at the bias in estimating mean uptake at different SBR settings. 
We conducted further analysis to understand this effect, where we looked at the distribution of the normalized error between the true and estimated uptake values over multiple experimental realizations. Results from this analysis (Fig.~\ref{fig:hist_error}) show that in a low-SBR setting, the estimated uptake in the circular regions were both below and above the true uptake in different experimental realizations whereas for higher SBR, the estimated activity was mostly below the true uptake. This resulted in a lower bias in low-SBR setting.

Our method was motivated by the goal of performing quantitative SPECT for $\alpha$-RPT. In this context, recently, a projection-domain quantification method, referred to as LC-QSPECT, has been proposed specifically for $\alpha$-RPT SPECT \cite{li2021projection}. The LC-QSPECT method estimates the activity in regions of interest directly from the projection data, skipping the reconstruction step. In contrast, the proposed method first performs a voxel-based reconstruction, followed by estimating activity from the reconstructed images. The LC-QSPECT method cannot represent heterogeneous structures within each region whereas the proposed method has the potential to represent such structures by virtue of being a reconstruction-based approach. Further, the proposed method operates directly with LM data, unlike the LC-QSPECT method that operates with binned data. Finally, the proposed method provides a mechanism to perform reconstruction in multi-energy window setting, while the LC-QSPECT method uses a single-energy window.

In this manuscript, we demonstrated the evaluation of the method in the context of imaging radium-223-based $\alpha$-RPT. However, the method is general and can be adapted to imaging other $\alpha$-particle emitting isotopes. Further, the method can also be applied to other SPECT isotopes, and is especially relevant for isotopes with multiple emission peaks. The method can also be used to process data from the scatter window, which may help improve performance on quantification tasks \cite{kadrmas1996svd,kadrmas1997analysis,rahman2020ieee}. 


In this manuscript, we advance upon the previously proposed path-based formalism to account for multiple emission peaks associated with applications including $\ra_223$-based $\alpha$-RPT SPECT. The path-based formalism enables developing the proposed reconstruction approach in the LM-MEW method while also explicitly accounting for the energy attribute. Moreover, this formalism accounts for image-degrading processes in SPECT including the attenuation and scatter of photons, and depth-dependent collimator response. The formalism can also be considered for other reconstruction, quantification, and image-quality evaluation methods, especially when using the energy attribute is desired.

While the theoretical formalism of the proposed method is in 3-D and can model any order of scatter, a main limitation of our evaluation study is that it was in 2-D and modeled only single-order of scatter. As mentioned earlier, these evaluations were proof-of-concept studies where our goal was to assess whether implementation of this method for more realistic 3-D settings would be beneficial. Our results provide supporting evidence for further development and evaluation of the reconstruction method to 3-D settings. A challenge in conducting these studies in 3-D is the high computational requirements. However, advances in parallel high-performance computing solutions provide a mechanism to address these challenges. In Appendix \ref{ap:comp_3d}, a detailed analysis of the computational requirements needed for advancing the evaluation study in 3-D multi-scatter setting is given. Another limitation of our evaluation is that our studies were conducted with digital phantoms. Advancing this method to 3-D will lay the platform to evaluate this method with physical phantom studies. Moreover, in our evaluation study, we considered six energy values for radium-223 isotope spanning from 81.1 keV to 270 keV. \rchange{More recent studies have shown that r}{R}adium-223 also has few higher emission peaks \rnew{\cite{kossert2015activity,pibida2017determination,collins2015precise}}. A future area of research is the inclusion of these higher-energy peaks while generating the LM data\rold{ \cite{kossert2015activity,pibida2017determination,collins2015precise}}. Additionally, for $\alpha$-RPT SPECT, stray-radiation-related noise is not negligible, given the low number of photon counts \cite{li2021projection}. Modeling this noise when performing the reconstruction operation with LM multi-energy-window data is another area of future research.

\section{Conclusion}
Towards addressing the challenge of quantitative SPECT for $\alpha$-particle radiopharmaceutical therapies, we propose a low-count list-mode SPECT reconstruction method for isotopes with multiple emission peaks. The method incorporates data from the multiple energy windows in LM format, including the energy attribute of each event.  The method was evaluated on the task of estimating mean activity within specific regions of interest in synthetic 2-D phantoms in a 2-D SPECT system within a single-scatter setup in the context of imaging the FDA-approved $\alpha$-particle radiopharmaceutical [$^{223}$Ra]RaCl${_2}$. The proposed method was also compared to methods where only single energy window was used and where the energy of the acquired data was binned. The results demonstrated that the proposed method that uses data in LM format, including the energy attribute, and includes data from multiple energy windows, yielded improved quantification performance compared to the use of more traditional single energy window or binned-data-based methods. Overall, the results provide promising evidence that using LM data, including the energy attribute for each event, and using data from multiple energy windows can improve performance on quantification tasks. These results motivate further development and validation of the method for low-count quantitative SPECT applications, including for $\alpha$-particle RPTs.
\clearpage

\section*{Appendix}
\section{Pseudo-code for LM-OSEM algorithm}
\label{ap:gpu_code}
 \removelatexerror
\begin{singlespace}
\begin{algorithm}[H]
\caption{Multi-GPU implementation of LM-OSEM algorithm
\label{GPU_algo}}
\DontPrintSemicolon
\SetAlgoLined
\SetAlgoNoLine
\SetKwInput{Input}{Input}
\SetKwInOut{Output}{Output}
\SetKw{Initialize}{Initialize}
\SetKw{PrecalculateStore}{Compute and Store}
\SetKw{Calculate}{Calculate}
\SetKw{Call}{Call}
\SetKw{cudaSetDevice}{cudaSetDevice}
\SetKw{Set}{Set}
\SetKwSty{textbf}
\SetFuncSty{texttt}
\SetKwFunction{FnSumAjKernel}{K\_SumAjKernel}
\SetKwFunction{FnSumZbarKernel}{K\_SumZbarKernel}
\SetKwFunction{FnAjActTermKernel}{K\_AjActTermKernel}
\SetKwFunction{FnAjAttnTermKernel}{K\_AjAttnTermKernel}
\SetKwFunction{FnFisherCalcKernel}{K\_CalcFisherKernel}
\SetKwFunction{CalcRadPath}{CalcRadPath}
\SetKwFunction{CalcDelPath}{CalcDelPath}
\SetKwData{sysParam}{sysParam}
\SetKwData{lmData}{lmData}
\SetKwData{}{}
\SetKwData{}{}

\Input{\;
\Indp
$\attn$ ($Q$-dimensional attenuation map), 
\lmData(Attributes of $J$ LM events),
$T$ (Acquisition time),
$numIter$ (\# main iterations),
$numSubIter$ (\#  sub-iterations)}
\KwData{\sysParam (SPECT system geometry)\;
}
\Output{$\f$ ($Q$-dimensional activity map)}
\nextnr\Initialize $\f\leftarrow\f^0$\;
\nextnr\PrecalculateStore {\;
\Indp
$\Delta_{q_m}(\subpath_{\r_u, k_v})$ for each subpath and voxel index\;
$\gamma(\r_u, \r_v, E_i)$ (Eq.~\eqref{eq:radpath}) \;
$s(\path)$ (Eq.~\eqref{eq:s_p}) for all paths \;
$N_{nz}$ (\# voxels with non-zero activity)  \;
}
\nextnr\Set $grid,~block,~shMem$ (grid, block and shared memory size, resp. for GPU kernel)\;
\nextnr\Set $nGPU$ (\# available GPUs) \;
\BlankLine
\nextnr\For{$i=0$ \KwTo $numIter-1$}{
\nextnr\For{$s=0$ \KwTo $numSubIter-1$}{
\nextnr\tcc{Compute and store $\sum_{w=1}^{W}\sum_{\path_q} \pr(\bAhat_j | \path_q,E_0^w) \lambda_{\mathcal{E}}(\path_q,E_0^w) s(\path_q,E_0^w)$ for each $j$ and $q$ in the array $sumAj$}
$sumAj[j,q] \leftarrow 0 \text{, for } j\in S_s^j; q=0,\ldots,N_{nz}-1$\;
\For{$q=0$ \KwTo $N_{nz}-1$}{
	\cudaSetDevice($q\%nGPU$)\;
	\Call{\FnSumAjKernel$<<<$grid,~block,~shMem$>>>(\f,~\attn, ~radPath,~voxIndex,~\boldsymbol{\Delta},~q,~s,$ \sysParam, \lmData, $ sumAj)$}
}

\BlankLine
\nextnr\tcc{Compute numerator of Eq.~\eqref{eq:os-lm-mlem}  $sumZbar[q]=\left\{ \sum_{j=1}^{J}\sum_{w=1}^{W}\sum_{\P_q}\bar{z}_{j,\P_q,w}^{(i+1)} \right\},q=0,1,...,Q-1$}
$sumZbar[q] \leftarrow \sum_{j\in S_s^j}\dfrac{sumAj[j,q]}{\sum_{q'} sumAj[j,q']}$\;
\nextnr\tcc{Update activity (Eq.~\eqref{eq:os-lm-mlem})}
\For{$q=0$ \KwTo $N_{nz}-1$}{
	$\lambda[q]\leftarrow \dfrac{sumZbar[q]}{T*\sum_{w=1}^{W}\sum_{\P_q \in S_s^\P} s(\P_q,E_0^w)}$
}
}
}
\end{algorithm}
\end{singlespace}

\section{Asymptotic analysis of the 2-D LM-OSEM algorithm}\label{ap:complexity}
In this section, we derive an expression for the computational complexity of the LM-OSEM algorithm (Eq.~\eqref{eq:os-lm-mlem}).
Denote the number of voxels in each slice as $Q$.
In a given iteration, denote the number of voxels with non-zero activity, the number of defined paths that start from voxel $q$ by $N_{nz}$ and $N_{\P}^q$ respectively.
Further, let $N_{\P,s}^q$ denote the number of defined paths that start from voxel $q$ and end in detector angles specified by subset $s$. Denote the number of events that belong to subset $s$ by $J_s^{2D}$. From Eqs.~\eqref{eq:os-lm-mlem} and \eqref{eq:z_bar_update_os}, we observe that each iteration requires approximately $\sum\limits_{q=1}^{N_{nz}} [ N_{\P,s}^{q}+(N_{\P,s}^q*J_s^{2D}) ]$ computations, assuming the cost of computing the denominator of Eq.~\eqref{eq:os-lm-mlem}) is negligible. 
Thus, the complexity of the $s^{\text{th}}$ iteration is given by $\mathcal{O}(N_{nz}J_s^{2D}\bar{N}_{\P,s})$, where $\bar{N}_{\P,s}$ is the mean of $N_{\P,s}^q$ over all voxels. 
Moreover, discretizing the 2-D orientation in $N_A$ samples and considering up to $L$ orders of scatters, we obtain $\bar{N}_{\P,s} = \mathcal{O}(N_A^{L}\sqrt{Q}^{L})$, where we used the fact that the computations related to the final scattering angles can be restricted to a constant set of angles because we know the detector angles for each event. 
Thus, the computational requirements of the  $s^{\text{th}}$ sub-iteration become $\mathcal{O}(N_{nz}J_s^{2D} N_A^{L}\sqrt{Q}^{L} )$.

\section{GPU implementation of the LM-OSEM algorithm}\label{ap:gpu_implementation}
Based on Eqs.~\eqref{eq:os-lm-mlem} \& \eqref{eq:z_bar_update_os}, at each iteration, we first compute the numerator of $\sum_{w=1}^{W}\sum_{\P_q}\bar{z}_{j,\P_q,w}$ for $j\in S_s^j$ and $q=0,\ldots N_{nz}-1$. Here, $N_{nz}$ denotes the number of voxels having non-zero initial activity.  To reduce computation time, the iteration was initialized with zero outside the estimated phantom boundary. The phantom mask estimation was done by a binned OSEM reconstruction without scatter modeling, and a morphological dilation and erosion operation. To calculate the numerator of $\sum_{w=1}^{W}\sum_{\P_q}\bar{z}_{j,\P_q,w}$ in parallel, the computation routine divided the computation of each voxel $q$ within an individual GPU. Inside each GPU, for a  fixed voxel $q_0$, the numerator of $\sum_{w=1}^{W}\sum_{\P_{q_0}}\bar{z}_{j,\P_{q_0},w}$ for each LM event $j$ in the subset was calculated in parallel. The denominator of $\sum_{w=1}^{W}\sum_{\P_{q}}\bar{z}_{j,\P_{q},w}$ is independent of $q$ and can be calculated by summing the numerators of $\sum_{w=1}^{W}\sum_{\P_q}\bar{z}_{j,\P_q,w}$ over all $q$.  Then, we updated the activity value using Eq.~\eqref{eq:os-lm-mlem}.

\section{Computational requirements for extending the LM-OSEM algorithm for 3-D}\label{ap:comp_3d}
In this section, we provide details about the computational requirements needed to extend the evaluation study in 3-D multi-scatter setting. For this purpose, we use the asymptotic analysis of 2-D LM-OSEM algorithm, described in Appendix \ref{ap:complexity}, as the starting point of the analysis. Denote $S$ as the number of slices. We assume that to discretize the orientation in 3-D, we use $N_A$ samples in azimuthal direction (each slice) and $N_E$ samples in elevation direction. The number of voxels will be $Q\times S$. \rchange{Thus, as per the treatment provided in the Appendix \ref{ap:complexity}, the computational requirement of the $s^{th}$ sub-iteration becomes $\mathcal{O}(S N_{nz} J_{s} N{_A}^{L} N_{E}^{L} (QS)^{(L/3)} )$. Thus, to advance the technique to 3-D in a single-scatter setting, the increase in computational requirement is of the order of $\mathcal{O}(S^{(4/3)} N_{E} )$, assuming that the number of list-mode events in each subset is the same in 2-D and 3-D case.}{Denote the number of events that belong to subset $s$ by $J_s^{3D}$. Thus, as per the treatment provided in the Appendix 8, the computational requirement of the $s^{\th}$ sub-iteration becomes $\mathcal{O}(S N_{nz} J_{s}^{3D} N{_A}^{L} N_{E}^{L} (QS)^{(L/3)} )$. Thus, to advance the technique to 3-D in a single-scatter setting, the increase in computational requirement is of the order of $\mathcal{O}(\frac{J_s^{3D}}{J_s^{2D}}S^{(4/3)} N_{E} )$.} \rchange{For example, assuming 64 slices and 128 discretizations in elevation direction, the computation time will increase by 32768 in the worst case. However, this calculation is based on the worst-case scenario, and}{However,} with careful optimization such as modeling scatter in lower resolution, we expect that the computation time could be decreased significantly. Thus, advancement in the computational hardware should enable evaluation in 3-D in a single-scatter setting. However, the addition of multi-scatter will increase the computation time exponentially, which may be challenging to handle by short-term advances in parallel computing. In this context, computationally-efficient methods have been developed previously to approximate multiple scatters for binned PET \cite{goggin1994model,ollinger1996model} and SPECT \cite{frey1996new}. We expect that algorithms that could accurately approximate multi-scatter settings in LM-SPECT may enable the implementation of this algorithm in 3-D multi-scatter setting with advances in parallel-computing hardware.
\section*{Acknowledgements}
This work was supported by the National Institute of Biomedical Imaging and Bioengineering of the National Institute of Health under grants R21-EB024647, R01 EB031051\rchange{ and }{, }R01- EB031962\rnew{, and NSF CAREER grant 2239707}. DLJT acknowledges support from grants NCI R01CA229893, R01CA240711, R01EB02925901, and the Siteman Cancer Center which is funded by NCI P30CA091842. Support is also acknowledged from the NVIDIA GPU Grant. We also thank the Scientific Compute Platform of Research Infrastructure Service (RIS) in Washington University for providing the computational resources.

\begin{backmatter}


\bibliographystyle{bmc-mathphys} 
\bibliography{ref}      


\begin{thebibliography}{47}
\ifx \bisbn   \undefined \def \bisbn  #1{ISBN #1}\fi
\ifx \binits  \undefined \def \binits#1{#1}\fi
\ifx \bauthor  \undefined \def \bauthor#1{#1}\fi
\ifx \batitle  \undefined \def \batitle#1{#1}\fi
\ifx \bjtitle  \undefined \def \bjtitle#1{#1}\fi
\ifx \bvolume  \undefined \def \bvolume#1{\textbf{#1}}\fi
\ifx \byear  \undefined \def \byear#1{#1}\fi
\ifx \bissue  \undefined \def \bissue#1{#1}\fi
\ifx \bfpage  \undefined \def \bfpage#1{#1}\fi
\ifx \blpage  \undefined \def \blpage #1{#1}\fi
\ifx \burl  \undefined \def \burl#1{\textsf{#1}}\fi
\ifx \doiurl  \undefined \def \doiurl#1{\textsf{#1}}\fi
\ifx \betal  \undefined \def \betal{\textit{et al.}}\fi
\ifx \binstitute  \undefined \def \binstitute#1{#1}\fi
\ifx \binstitutionaled  \undefined \def \binstitutionaled#1{#1}\fi
\ifx \bctitle  \undefined \def \bctitle#1{#1}\fi
\ifx \beditor  \undefined \def \beditor#1{#1}\fi
\ifx \bpublisher  \undefined \def \bpublisher#1{#1}\fi
\ifx \bbtitle  \undefined \def \bbtitle#1{#1}\fi
\ifx \bedition  \undefined \def \bedition#1{#1}\fi
\ifx \bseriesno  \undefined \def \bseriesno#1{#1}\fi
\ifx \blocation  \undefined \def \blocation#1{#1}\fi
\ifx \bsertitle  \undefined \def \bsertitle#1{#1}\fi
\ifx \bsnm \undefined \def \bsnm#1{#1}\fi
\ifx \bsuffix \undefined \def \bsuffix#1{#1}\fi
\ifx \bparticle \undefined \def \bparticle#1{#1}\fi
\ifx \barticle \undefined \def \barticle#1{#1}\fi
\ifx \bconfdate \undefined \def \bconfdate #1{#1}\fi
\ifx \botherref \undefined \def \botherref #1{#1}\fi
\ifx \url \undefined \def \url#1{\textsf{#1}}\fi
\ifx \bchapter \undefined \def \bchapter#1{#1}\fi
\ifx \bbook \undefined \def \bbook#1{#1}\fi
\ifx \bcomment \undefined \def \bcomment#1{#1}\fi
\ifx \oauthor \undefined \def \oauthor#1{#1}\fi
\ifx \citeauthoryear \undefined \def \citeauthoryear#1{#1}\fi
\ifx \endbibitem  \undefined \def \endbibitem {}\fi
\ifx \bconflocation  \undefined \def \bconflocation#1{#1}\fi
\ifx \arxivurl  \undefined \def \arxivurl#1{\textsf{#1}}\fi
\csname PreBibitemsHook\endcsname

\bibitem{abou2020prostate}
\begin{barticle}
\bauthor{\bsnm{Abou}, \binits{D.}},
\bauthor{\bsnm{Benabdallah}, \binits{N.}},
\bauthor{\bsnm{Jiang}, \binits{W.}},
\bauthor{\bsnm{Peng}, \binits{L.}},
\bauthor{\bsnm{Zhang}, \binits{H.}},
\bauthor{\bsnm{Villmer}, \binits{A.}},
\bauthor{\bsnm{Longtine}, \binits{M.S.}},
\bauthor{\bsnm{Thorek}, \binits{D.L.}}:
\batitle{{Prostate cancer theranostics-an overview}}.
\bjtitle{Front. Oncol.}
\bvolume{10},
\bfpage{884}
(\byear{2020})
\end{barticle}
\endbibitem

\bibitem{baidoo2013molecular}
\begin{barticle}
\bauthor{\bsnm{Baidoo}, \binits{K.E.}},
\bauthor{\bsnm{Yong}, \binits{K.}},
\bauthor{\bsnm{Brechbiel}, \binits{M.W.}}:
\batitle{{Molecular Pathways: Targeted $\alpha$-Particle Radiation
  TherapyTargeted $\alpha$-Particle Radiation Therapy Mechanisms}}.
\bjtitle{Clinical cancer research}
\bvolume{19}(\bissue{3}),
\bfpage{530}--\blpage{537}
(\byear{2013})
\end{barticle}
\endbibitem

\bibitem{kluetz2014radium}
\begin{barticle}
\bauthor{\bsnm{Kluetz}, \binits{P.G.}},
\bauthor{\bsnm{Pierce}, \binits{W.}},
\bauthor{\bsnm{Maher}, \binits{V.E.}},
\bauthor{\bsnm{Zhang}, \binits{H.}},
\bauthor{\bsnm{Tang}, \binits{S.}},
\bauthor{\bsnm{Song}, \binits{P.}},
\bauthor{\bsnm{Liu}, \binits{Q.}},
\bauthor{\bsnm{Haber}, \binits{M.T.}},
\bauthor{\bsnm{Leutzinger}, \binits{E.E.}},
\bauthor{\bsnm{Al-Hakim}, \binits{A.}}, \betal:
\batitle{{Radium Ra 223 dichloride injection: US Food and Drug Administration
  drug approval summary}}.
\bjtitle{Clin. Cancer Res.}
\bvolume{20}(\bissue{1}),
\bfpage{9}--\blpage{14}
(\byear{2014})
\end{barticle}
\endbibitem

\bibitem{kratochwil2017targeted}
\begin{barticle}
\bauthor{\bsnm{Kratochwil}, \binits{C.}},
\bauthor{\bsnm{Bruchertseifer}, \binits{F.}},
\bauthor{\bsnm{Rathke}, \binits{H.}},
\bauthor{\bsnm{Bronzel}, \binits{M.}},
\bauthor{\bsnm{Apostolidis}, \binits{C.}},
\bauthor{\bsnm{Weichert}, \binits{W.}},
\bauthor{\bsnm{Haberkorn}, \binits{U.}},
\bauthor{\bsnm{Giesel}, \binits{F.L.}},
\bauthor{\bsnm{Morgenstern}, \binits{A.}}:
\batitle{{Targeted alpha-therapy of metastatic castration-resistant prostate
  cancer with 225Ac-PSMA-617: dosimetry estimate and empiric dose finding}}.
\bjtitle{J. Nucl. Med.}
\bvolume{58}(\bissue{10}),
\bfpage{1624}--\blpage{1631}
(\byear{2017})
\end{barticle}
\endbibitem

\bibitem{larsson2020feasibility}
\begin{barticle}
\bauthor{\bsnm{Larsson}, \binits{E.}},
\bauthor{\bsnm{Brolin}, \binits{G.}},
\bauthor{\bsnm{Cleton}, \binits{A.}},
\bauthor{\bsnm{Ohlsson}, \binits{T.}},
\bauthor{\bsnm{Lind{\'e}n}, \binits{O.}},
\bauthor{\bsnm{Hindorf}, \binits{C.}}:
\batitle{{Feasibility of thorium-227/radium-223 gamma-camera imaging during
  radionuclide therapy}}.
\bjtitle{Cancer Biother. Radiopharm.}
\bvolume{35}(\bissue{7}),
\bfpage{540}--\blpage{548}
(\byear{2020})
\end{barticle}
\endbibitem

\bibitem{murray2020quantitative}
\begin{barticle}
\bauthor{\bsnm{Murray}, \binits{I.}},
\bauthor{\bsnm{Rojas}, \binits{B.}},
\bauthor{\bsnm{Gear}, \binits{J.}},
\bauthor{\bsnm{Callister}, \binits{R.}},
\bauthor{\bsnm{Cleton}, \binits{A.}},
\bauthor{\bsnm{Flux}, \binits{G.D.}}:
\batitle{{Quantitative Dual-Isotope Planar Imaging of Thorium-227 and
  Radium-223 Using Defined Energy Windows}}.
\bjtitle{Cancer Biother. Radiopharm.}
\bvolume{35}(\bissue{7}),
\bfpage{530}--\blpage{539}
(\byear{2020})
\end{barticle}
\endbibitem

\bibitem{ghaly2019quantitative}
\begin{botherref}
\oauthor{\bsnm{Ghaly}, \binits{M.}},
\oauthor{\bsnm{Sgouros}, \binits{G.}},
\oauthor{\bsnm{Frey}, \binits{E.}}:
{Quantitative dual isotope SPECT imaging of the alpha-emitters Th-227 and
  Ra-223}.
Soc Nuclear Med
(2019)
\end{botherref}
\endbibitem

\bibitem{gustafsson2020feasibility}
\begin{barticle}
\bauthor{\bsnm{Gustafsson}, \binits{J.}},
\bauthor{\bsnm{Rode{\~n}o}, \binits{E.}},
\bauthor{\bsnm{M{\'\i}nguez}, \binits{P.}}:
\batitle{{Feasibility and limitations of quantitative SPECT for 223Ra}}.
\bjtitle{Phys. Med. Biol.}
\bvolume{65}(\bissue{8}),
\bfpage{085012}
(\byear{2020})
\end{barticle}
\endbibitem

\bibitem{tafreshi2019development}
\begin{barticle}
\bauthor{\bsnm{Tafreshi}, \binits{N.K.}},
\bauthor{\bsnm{Doligalski}, \binits{M.L.}},
\bauthor{\bsnm{Tichacek}, \binits{C.J.}},
\bauthor{\bsnm{Pandya}, \binits{D.N.}},
\bauthor{\bsnm{Budzevich}, \binits{M.M.}},
\bauthor{\bsnm{El-Haddad}, \binits{G.}},
\bauthor{\bsnm{Khushalani}, \binits{N.I.}},
\bauthor{\bsnm{Moros}, \binits{E.G.}},
\bauthor{\bsnm{McLaughlin}, \binits{M.L.}},
\bauthor{\bsnm{Wadas}, \binits{T.J.}}, \betal:
\batitle{{Development of targeted alpha particle therapy for solid tumors}}.
\bjtitle{Molecules}
\bvolume{24}(\bissue{23}),
\bfpage{4314}
(\byear{2019})
\end{barticle}
\endbibitem

\bibitem{brans2007clinical}
\begin{barticle}
\bauthor{\bsnm{Brans}, \binits{B.}},
\bauthor{\bsnm{Bodei}, \binits{L.}},
\bauthor{\bsnm{Giammarile}, \binits{F.}},
\bauthor{\bsnm{Lind{\'e}n}, \binits{O.}},
\bauthor{\bsnm{Luster}, \binits{M.}},
\bauthor{\bsnm{Oyen}, \binits{W.}},
\bauthor{\bsnm{Tennvall}, \binits{J.}}:
\batitle{Clinical radionuclide therapy dosimetry: the quest for the “holy
  gray”}.
\bjtitle{Eur. J. Nucl. Med. Mol. Imaging}
\bvolume{34}(\bissue{5}),
\bfpage{772}--\blpage{786}
(\byear{2007})
\end{barticle}
\endbibitem

\bibitem{benabdallah2019223ra}
\begin{barticle}
\bauthor{\bsnm{Benabdallah}, \binits{N.}},
\bauthor{\bsnm{Bernardini}, \binits{M.}},
\bauthor{\bsnm{Bianciardi}, \binits{M.}},
\bauthor{\bparticle{de} \bsnm{Labriolle-Vaylet}, \binits{C.}},
\bauthor{\bsnm{Franck}, \binits{D.}},
\bauthor{\bsnm{Desbr{\'e}e}, \binits{A.}}:
\batitle{{223Ra-dichloride therapy of bone metastasis: optimization of SPECT
  images for quantification}}.
\bjtitle{EJNMMI research}
\bvolume{9}(\bissue{1}),
\bfpage{1}--\blpage{12}
(\byear{2019})
\end{barticle}
\endbibitem

\bibitem{osti_22632144}
\begin{botherref}
\oauthor{\bsnm{Yue}, \binits{J.}},
\oauthor{\bsnm{Hobbs}, \binits{R.}},
\oauthor{\bsnm{Sgouros}, \binits{G.}},
\oauthor{\bsnm{Frey}, \binits{E.}}:
{SU-F-J-08: Quantitative SPECT Imaging of Ra-223 in a Phantom}.
Med. Phys.
\textbf{43}(6)
(2016).
doi:\doiurl{10.1118/1.4955916}
\end{botherref}
\endbibitem

\bibitem{li2021projection}
\begin{botherref}
\oauthor{\bsnm{Li}, \binits{Z.}},
\oauthor{\bsnm{Benabdallah}, \binits{N.}},
\oauthor{\bsnm{Abou}, \binits{D.S.}},
\oauthor{\bsnm{Baumann}, \binits{B.C.}},
\oauthor{\bsnm{Dehdashti}, \binits{F.}},
\oauthor{\bsnm{Ballard}, \binits{D.H.}},
\oauthor{\bsnm{Liu}, \binits{J.}},
\oauthor{\bsnm{Jammalamadaka}, \binits{U.}},
\oauthor{\bsnm{Laforest}, \binits{R.}},
\oauthor{\bsnm{Wahl}, \binits{R.L.}},
\oauthor{\bsnm{Thorek}, \binits{D.L.J.}},
\oauthor{\bsnm{Jha}, \binits{A.K.}}:
{A projection-domain low-count quantitative SPECT method for $\alpha$-particle
  emitting radiopharmaceutical therapy}.
IEEE Trans. Radiat. Plasma Med. Sci.,
1--1
(2022).
doi:\doiurl{10.1109/TRPMS.2022.3175435}
\end{botherref}
\endbibitem

\bibitem{abou2020preclinical}
\begin{barticle}
\bauthor{\bsnm{Abou}, \binits{D.S.}},
\bauthor{\bsnm{Rittenbach}, \binits{A.}},
\bauthor{\bsnm{Tomlinson}, \binits{R.E.}},
\bauthor{\bsnm{Finley}, \binits{P.A.}},
\bauthor{\bsnm{Tsui}, \binits{B.}},
\bauthor{\bsnm{Simons}, \binits{B.W.}},
\bauthor{\bsnm{Jha}, \binits{A.K.}},
\bauthor{\bsnm{Ulmert}, \binits{D.}},
\bauthor{\bsnm{Riddle}, \binits{R.C.}},
\bauthor{\bsnm{Thorek}, \binits{D.L.}}:
\batitle{{Preclinical single photon emission computed tomography of alpha
  particle-emitting radium-223}}.
\bjtitle{Cancer Biother. Radiopharm.}
\bvolume{35}(\bissue{7}),
\bfpage{520}--\blpage{529}
(\byear{2020})
\end{barticle}
\endbibitem

\bibitem{owaki2017ra}
\begin{barticle}
\bauthor{\bsnm{Owaki}, \binits{Y.}},
\bauthor{\bsnm{Nakahara}, \binits{T.}},
\bauthor{\bsnm{Kosaka}, \binits{T.}},
\bauthor{\bsnm{Fukada}, \binits{J.}},
\bauthor{\bsnm{Kumabe}, \binits{A.}},
\bauthor{\bsnm{Ichimura}, \binits{A.}},
\bauthor{\bsnm{Murakami}, \binits{M.}},
\bauthor{\bsnm{Nakajima}, \binits{K.}},
\bauthor{\bsnm{Fukushi}, \binits{M.}},
\bauthor{\bsnm{Inoue}, \binits{K.}}, \betal:
\batitle{{Ra-223 SPECT for semi-quantitative analysis in comparison with Tc-99m
  HMDP SPECT: phantom study and initial clinical experience}}.
\bjtitle{EJNMMI research}
\bvolume{7}(\bissue{1}),
\bfpage{1}--\blpage{11}
(\byear{2017})
\end{barticle}
\endbibitem

\bibitem{benabdallah2021practical}
\begin{barticle}
\bauthor{\bsnm{Benabdallah}, \binits{N.}},
\bauthor{\bsnm{Scheve}, \binits{W.}},
\bauthor{\bsnm{Dunn}, \binits{N.}},
\bauthor{\bsnm{Silvestros}, \binits{D.}},
\bauthor{\bsnm{Schelker}, \binits{P.}},
\bauthor{\bsnm{Abou}, \binits{D.}},
\bauthor{\bsnm{Jammalamadaka}, \binits{U.}},
\bauthor{\bsnm{Laforest}, \binits{R.}},
\bauthor{\bsnm{Li}, \binits{Z.}},
\bauthor{\bsnm{Liu}, \binits{J.}}, \betal:
\batitle{{Practical considerations for quantitative clinical SPECT/CT imaging
  of alpha particle emitting radioisotopes}}.
\bjtitle{Theranostics}
\bvolume{11}(\bissue{20}),
\bfpage{9721}
(\byear{2021})
\end{barticle}
\endbibitem

\bibitem{chun2019algorithms}
\begin{barticle}
\bauthor{\bsnm{Chun}, \binits{S.Y.}},
\bauthor{\bsnm{Nguyen}, \binits{M.P.}},
\bauthor{\bsnm{Phan}, \binits{T.Q.}},
\bauthor{\bsnm{Kim}, \binits{H.}},
\bauthor{\bsnm{Fessler}, \binits{J.A.}},
\bauthor{\bsnm{Dewaraja}, \binits{Y.K.}}:
\batitle{{Algorithms and analyses for joint spectral image reconstruction in
  Y-90 bremsstrahlung SPECT}}.
\bjtitle{IEEE Trans. Med. Imaging}
\bvolume{39}(\bissue{5}),
\bfpage{1369}--\blpage{1379}
(\byear{2019})
\end{barticle}
\endbibitem

\bibitem{kadrmas1996svd}
\begin{barticle}
\bauthor{\bsnm{Kadrmas}, \binits{D.J.}},
\bauthor{\bsnm{Frey}, \binits{E.C.}},
\bauthor{\bsnm{Tsui}, \binits{B.M.}}:
\batitle{{An SVD investigation of modeling scatter in multiple energy windows
  for improved SPECT images}}.
\bjtitle{IEEE Trans. Nucl. Sci.}
\bvolume{43}(\bissue{4}),
\bfpage{2275}--\blpage{2284}
(\byear{1996})
\end{barticle}
\endbibitem

\bibitem{kadrmas1997analysis}
\begin{barticle}
\bauthor{\bsnm{Kadrmas}, \binits{D.J.}},
\bauthor{\bsnm{Frey}, \binits{E.C.}},
\bauthor{\bsnm{Tsui}, \binits{B.M.}}:
\batitle{Analysis of the reconstructibility and noise properties of scattered
  photons in {T}c-99m {SPECT}}.
\bjtitle{Phys. Med. Biol.}
\bvolume{42}(\bissue{12}),
\bfpage{2493}
(\byear{1997})
\end{barticle}
\endbibitem

\bibitem{clarkson2020quantifying}
\begin{barticle}
\bauthor{\bsnm{Clarkson}, \binits{E.}},
\bauthor{\bsnm{Kupinski}, \binits{M.}}:
\batitle{Quantifying the loss of information from binning list-mode data}.
\bjtitle{JOSA A}
\bvolume{37}(\bissue{3}),
\bfpage{450}--\blpage{457}
(\byear{2020})
\end{barticle}
\endbibitem

\bibitem{jha2015estimating}
\begin{bchapter}
\bauthor{\bsnm{Jha}, \binits{A.K.}},
\bauthor{\bsnm{Frey}, \binits{E.C.}}:
\bctitle{Estimating {ROI} activity concentration with photon-processing and
  photon-counting {SPECT} imaging systems}.
In: \bbtitle{Proc. SPIE Med. Imag.},
vol. \bseriesno{9412},
p. \bfpage{94120}
(\byear{2015}).
\bcomment{International Society for Optics and Photonics}
\end{bchapter}
\endbibitem

\bibitem{jha2015singular}
\begin{barticle}
\bauthor{\bsnm{Jha}, \binits{A.K.}},
\bauthor{\bsnm{Barrett}, \binits{H.H.}},
\bauthor{\bsnm{Frey}, \binits{E.C.}},
\bauthor{\bsnm{Clarkson}, \binits{E.}},
\bauthor{\bsnm{Caucci}, \binits{L.}},
\bauthor{\bsnm{Kupinski}, \binits{M.A.}}:
\batitle{Singular value decomposition for photon-processing nuclear imaging
  systems and applications for reconstruction and computing null functions}.
\bjtitle{Phys. Med. Biol.}
\bvolume{60}(\bissue{18}),
\bfpage{7359}
(\byear{2015})
\end{barticle}
\endbibitem

\bibitem{caucci2019towards}
\begin{barticle}
\bauthor{\bsnm{Caucci}, \binits{L.}},
\bauthor{\bsnm{Liu}, \binits{Z.}},
\bauthor{\bsnm{Jha}, \binits{A.}},
\bauthor{\bsnm{Han}, \binits{H.}},
\bauthor{\bsnm{Furenlid}, \binits{L.}},
\bauthor{\bsnm{Barrett}, \binits{H.}}:
\batitle{Towards continuous-to-continuous 3{D} imaging in the real world}.
\bjtitle{Phys. Med. Biol.}
\bvolume{64}(\bissue{18}),
\bfpage{185007}
(\byear{2019})
\end{barticle}
\endbibitem

\bibitem{rahman2020ieee}
\begin{bchapter}
\bauthor{\bsnm{{Rahman}}, \binits{M.A.}},
\bauthor{\bsnm{{Laforest}}, \binits{R.}},
\bauthor{\bsnm{{Jha}}, \binits{A.K.}}:
\bctitle{A list-mode {OSEM}-based attenuation and scatter compensation method
  for {SPECT}}.
In: \bbtitle{2020 IEEE 17th International Symposium on Biomedical Imaging
  (ISBI)},
pp. \bfpage{646}--\blpage{650}
(\byear{2020}).
doi:\doiurl{10.1109/ISBI45749.2020.9098333}
\end{bchapter}
\endbibitem

\bibitem{bousse2016joint}
\begin{bchapter}
\bauthor{\bsnm{Bousse}, \binits{A.}},
\bauthor{\bsnm{Sidlesky}, \binits{A.}},
\bauthor{\bsnm{Roth}, \binits{N.}},
\bauthor{\bsnm{Rashidnasab}, \binits{A.}},
\bauthor{\bsnm{Thielemans}, \binits{K.}},
\bauthor{\bsnm{Hutton}, \binits{B.F.}}:
\bctitle{Joint activity/attenuation reconstruction in spect using photopeak and
  scatter sinograms}.
In: \bbtitle{2016 IEEE Nuclear Science Symposium, Medical Imaging Conference
  and Room-Temperature Semiconductor Detector Workshop (NSS/MIC/RTSD)},
pp. \bfpage{1}--\blpage{4}
(\byear{2016}).
\bcomment{IEEE}
\end{bchapter}
\endbibitem

\bibitem{arridge2021overview}
\begin{barticle}
\bauthor{\bsnm{Arridge}, \binits{S.R.}},
\bauthor{\bsnm{Ehrhardt}, \binits{M.J.}},
\bauthor{\bsnm{Thielemans}, \binits{K.}}:
\batitle{(an overview of) synergistic reconstruction for
  multimodality/multichannel imaging methods}.
\bjtitle{Philosophical Transactions of the Royal Society A}
\bvolume{379}(\bissue{2200}),
\bfpage{20200205}
(\byear{2021})
\end{barticle}
\endbibitem

\bibitem{rahman2020fisher}
\begin{barticle}
\bauthor{\bsnm{Rahman}, \binits{M.A.}},
\bauthor{\bsnm{Zhu}, \binits{Y.}},
\bauthor{\bsnm{Clarkson}, \binits{E.}},
\bauthor{\bsnm{Kupinski}, \binits{M.A.}},
\bauthor{\bsnm{Frey}, \binits{E.C.}},
\bauthor{\bsnm{Jha}, \binits{A.K.}}:
\batitle{Fisher information analysis of list-mode spect emission data for joint
  estimation of activity and attenuation distribution}.
\bjtitle{Inverse Problems}
\bvolume{36}(\bissue{8}),
\bfpage{084002}
(\byear{2020})
\end{barticle}
\endbibitem

\bibitem{guerin2010novel}
\begin{barticle}
\bauthor{\bsnm{Gu{\'e}rin}, \binits{B.}},
\bauthor{\bsnm{El~Fakhri}, \binits{G.}}:
\batitle{Novel scatter compensation of list-mode {PET} data using spatial and
  energy dependent corrections}.
\bjtitle{IEEE Trans. Med. Imaging}
\bvolume{30}(\bissue{3}),
\bfpage{759}--\blpage{773}
(\byear{2010})
\end{barticle}
\endbibitem

\bibitem{jha2013retrieving}
\begin{botherref}
\oauthor{\bsnm{Jha}, \binits{A.K.}}:
Retrieving information from scattered photons in medical imaging
(2013)
\end{botherref}
\endbibitem

\bibitem{jha2013joint}
\begin{bchapter}
\bauthor{\bsnm{Jha}, \binits{A.K.}},
\bauthor{\bsnm{Clarkson}, \binits{E.}},
\bauthor{\bsnm{Kupinski}, \binits{M.A.}},
\bauthor{\bsnm{Barrett}, \binits{H.H.}}:
\bctitle{Joint reconstruction of activity and attenuation map using {LM}
  {SPECT} emission data}.
In: \bbtitle{Proc. SPIE Med. Imag.},
vol. \bseriesno{8668},
p. \bfpage{86681}
(\byear{2013}).
\bcomment{International Society for Optics and Photonics}
\end{bchapter}
\endbibitem

\bibitem{barrett1997list}
\begin{barticle}
\bauthor{\bsnm{Barrett}, \binits{H.H.}},
\bauthor{\bsnm{White}, \binits{T.}},
\bauthor{\bsnm{Parra}, \binits{L.C.}}:
\batitle{{List-mode likelihood}}.
\bjtitle{JOSA A}
\bvolume{14}(\bissue{11}),
\bfpage{2914}--\blpage{2923}
(\byear{1997})
\end{barticle}
\endbibitem

\bibitem{klein1929streuung}
\begin{barticle}
\bauthor{\bsnm{Klein}, \binits{O.}},
\bauthor{\bsnm{Nishina}, \binits{Y.}}:
\batitle{{{\"U}ber die Streuung von Strahlung durch freie Elektronen nach der
  neuen relativistischen Quantendynamik von Dirac}}.
\bjtitle{Zeitschrift f{\"u}r Physik}
\bvolume{52}(\bissue{11}),
\bfpage{853}--\blpage{868}
(\byear{1929})
\end{barticle}
\endbibitem

\bibitem{shepp1982maximum}
\begin{barticle}
\bauthor{\bsnm{Shepp}, \binits{L.A.}},
\bauthor{\bsnm{Vardi}, \binits{Y.}}:
\batitle{{Maximum likelihood reconstruction for emission tomography}}.
\bjtitle{IEEE Trans. Med. Imaging}
\bvolume{1}(\bissue{2}),
\bfpage{113}--\blpage{122}
(\byear{1982})
\end{barticle}
\endbibitem

\bibitem{lange1984reconstruction}
\begin{barticle}
\bauthor{\bsnm{Lange}, \binits{K.}},
\bauthor{\bsnm{Carson}, \binits{R.}}, \betal:
\batitle{Em reconstruction algorithms for emission and transmission
  tomography}.
\bjtitle{J. Comput. Assist. Tomogr.}
\bvolume{8}(\bissue{2}),
\bfpage{306}--\blpage{16}
(\byear{1984})
\end{barticle}
\endbibitem

\bibitem{parra1998list}
\begin{barticle}
\bauthor{\bsnm{Parra}, \binits{L.}},
\bauthor{\bsnm{Barrett}, \binits{H.H.}}:
\batitle{{List-mode likelihood: EM algorithm and image quality estimation
  demonstrated on 2-D PET}}.
\bjtitle{IEEE Trans. Med. Imaging}
\bvolume{17}(\bissue{2}),
\bfpage{228}--\blpage{235}
(\byear{1998})
\end{barticle}
\endbibitem

\bibitem{khurd2004globally}
\begin{barticle}
\bauthor{\bsnm{Khurd}, \binits{P.}},
\bauthor{\bsnm{Hsiao}, \binits{T.}},
\bauthor{\bsnm{Rangarajan}, \binits{A.}},
\bauthor{\bsnm{Gindi}, \binits{G.}}:
\batitle{{A globally convergent regularized ordered-subset EM algorithm for
  list-mode reconstruction}}.
\bjtitle{IEEE Trans. Nucl. Sci.}
\bvolume{51}(\bissue{3}),
\bfpage{719}--\blpage{725}
(\byear{2004})
\end{barticle}
\endbibitem

\bibitem{hudson1994accelerated}
\begin{barticle}
\bauthor{\bsnm{Hudson}, \binits{H.M.}},
\bauthor{\bsnm{Larkin}, \binits{R.S.}}:
\batitle{{Accelerated image reconstruction using ordered subsets of projection
  data}}.
\bjtitle{IEEE Trans. Med. Imaging}
\bvolume{13}(\bissue{4}),
\bfpage{601}--\blpage{609}
(\byear{1994})
\end{barticle}
\endbibitem

\bibitem{bochud1999statistical}
\begin{barticle}
\bauthor{\bsnm{Bochud}, \binits{F.O.}},
\bauthor{\bsnm{Abbey}, \binits{C.K.}},
\bauthor{\bsnm{Eckstein}, \binits{M.P.}}:
\batitle{Statistical texture synthesis of mammographic images with clustered
  lumpy backgrounds}.
\bjtitle{Opt. Express}
\bvolume{4}(\bissue{1}),
\bfpage{33}--\blpage{43}
(\byear{1999})
\end{barticle}
\endbibitem

\bibitem{hindorf2012quantitative}
\begin{barticle}
\bauthor{\bsnm{Hindorf}, \binits{C.}},
\bauthor{\bsnm{Chittenden}, \binits{S.}},
\bauthor{\bsnm{Aksnes}, \binits{A.-K.}},
\bauthor{\bsnm{Parker}, \binits{C.}},
\bauthor{\bsnm{Flux}, \binits{G.D.}}:
\batitle{{Quantitative imaging of 223Ra-chloride (Alpharadin) for targeted
  alpha-emitting radionuclide therapy of bone metastases}}.
\bjtitle{Nucl. Med. Commun.}
\bvolume{33}(\bissue{7}),
\bfpage{726}--\blpage{732}
(\byear{2012})
\end{barticle}
\endbibitem

\bibitem{prince2006medical}
\begin{bbook}
\bauthor{\bsnm{Prince}, \binits{J.L.}},
\bauthor{\bsnm{Links}, \binits{J.M.}}:
\bbtitle{Medical Imaging Signals and Systems}
vol. \bseriesno{37}.
\bpublisher{Pearson Prentice Hall Upper Saddle River}, \blocation{???}
(\byear{2006})
\end{bbook}
\endbibitem

\bibitem{cloquet2011does}
\begin{bchapter}
\bauthor{\bsnm{Cloquet}, \binits{C.}},
\bauthor{\bsnm{Defrise}, \binits{M.}}:
\bctitle{{Does OSEM achieve the lowest variance?}}
In: \bbtitle{2011 IEEE Nuclear Science Symposium Conference Record},
pp. \bfpage{2360}--\blpage{2365}
(\byear{2011}).
\bcomment{IEEE}
\end{bchapter}
\endbibitem

\bibitem{kossert2015activity}
\begin{barticle}
\bauthor{\bsnm{Kossert}, \binits{K.}},
\bauthor{\bsnm{Bokeloh}, \binits{K.}},
\bauthor{\bsnm{Dersch}, \binits{R.}},
\bauthor{\bsnm{N{\"a}hle}, \binits{O.}}:
\batitle{{Activity determination of 227Ac and 223Ra by means of liquid
  scintillation counting and determination of nuclear decay data}}.
\bjtitle{Appl. Radiat. Isot.}
\bvolume{95},
\bfpage{143}--\blpage{152}
(\byear{2015})
\end{barticle}
\endbibitem

\bibitem{pibida2017determination}
\begin{barticle}
\bauthor{\bsnm{Pibida}, \binits{L.}},
\bauthor{\bsnm{Zimmerman}, \binits{B.}},
\bauthor{\bsnm{Bergeron}, \binits{D.E.}},
\bauthor{\bsnm{Fitzgerald}, \binits{R.}},
\bauthor{\bsnm{Cessna}, \binits{J.T.}},
\bauthor{\bsnm{King}, \binits{L.}}:
\batitle{{Determination of photon emission probability for the main gamma ray
  and half-life measurements of 64Cu}}.
\bjtitle{Appl. Radiat. Isot.}
\bvolume{129},
\bfpage{6}--\blpage{12}
(\byear{2017})
\end{barticle}
\endbibitem

\bibitem{collins2015precise}
\begin{barticle}
\bauthor{\bsnm{Collins}, \binits{S.}},
\bauthor{\bsnm{Pearce}, \binits{A.}},
\bauthor{\bsnm{Regan}, \binits{P.}},
\bauthor{\bsnm{Keightley}, \binits{J.}}:
\batitle{{Precise measurements of the absolute $\gamma$-ray emission
  probabilities of 223Ra and decay progeny in equilibrium}}.
\bjtitle{Appl. Radiat. Isot.}
\bvolume{102},
\bfpage{15}--\blpage{28}
(\byear{2015})
\end{barticle}
\endbibitem

\bibitem{goggin1994model}
\begin{bchapter}
\bauthor{\bsnm{Goggin}, \binits{A.S.}},
\bauthor{\bsnm{Ollinger}, \binits{J.M.}}:
\bctitle{{A model for multiple scatters in fully 3D PET}}.
In: \bbtitle{Proceedings of 1994 IEEE Nuclear Science Symposium-NSS'94},
vol. \bseriesno{4},
pp. \bfpage{1609}--\blpage{1613}
(\byear{1994}).
\bcomment{IEEE}
\end{bchapter}
\endbibitem

\bibitem{ollinger1996model}
\begin{barticle}
\bauthor{\bsnm{Ollinger}, \binits{J.M.}}:
\batitle{{Model-based scatter correction for fully 3D PET}}.
\bjtitle{Phys. Med. Biol.}
\bvolume{41}(\bissue{1}),
\bfpage{153}
(\byear{1996})
\end{barticle}
\endbibitem

\bibitem{frey1996new}
\begin{bchapter}
\bauthor{\bsnm{Frey}, \binits{E.C.}},
\bauthor{\bsnm{Tsui}, \binits{B.}}:
\bctitle{{A new method for modeling the spatially-variant, object-dependent
  scatter response function in SPECT}}.
In: \bbtitle{1996 IEEE Nuclear Science Symposium. Conference Record},
vol. \bseriesno{2},
pp. \bfpage{1082}--\blpage{1086}
(\byear{1996}).
\bcomment{IEEE}
\end{bchapter}
\endbibitem

\end{thebibliography}

\newcommand{\BMCxmlcomment}[1]{}

\BMCxmlcomment{

<refgrp>

<bibl id="B1">
  <title><p>{Prostate cancer theranostics-an overview}</p></title>
  <aug>
    <au><snm>Abou</snm><fnm>D</fnm></au>
    <au><snm>Benabdallah</snm><fnm>N</fnm></au>
    <au><snm>Jiang</snm><fnm>W</fnm></au>
    <au><snm>Peng</snm><fnm>L</fnm></au>
    <au><snm>Zhang</snm><fnm>H</fnm></au>
    <au><snm>Villmer</snm><fnm>A</fnm></au>
    <au><snm>Longtine</snm><fnm>MS</fnm></au>
    <au><snm>Thorek</snm><fnm>DL</fnm></au>
  </aug>
  <source>Front. Oncol.</source>
  <publisher>Frontiers Media SA</publisher>
  <pubdate>2020</pubdate>
  <volume>10</volume>
  <fpage>884</fpage>
</bibl>

<bibl id="B2">
  <title><p>{Molecular Pathways: Targeted $\alpha$-Particle Radiation
  TherapyTargeted $\alpha$-Particle Radiation Therapy Mechanisms}</p></title>
  <aug>
    <au><snm>Baidoo</snm><fnm>KE</fnm></au>
    <au><snm>Yong</snm><fnm>K</fnm></au>
    <au><snm>Brechbiel</snm><fnm>MW</fnm></au>
  </aug>
  <source>Clinical cancer research</source>
  <publisher>AACR</publisher>
  <pubdate>2013</pubdate>
  <volume>19</volume>
  <issue>3</issue>
  <fpage>530</fpage>
  <lpage>-537</lpage>
</bibl>

<bibl id="B3">
  <title><p>{Radium Ra 223 dichloride injection: US Food and Drug
  Administration drug approval summary}</p></title>
  <aug>
    <au><snm>Kluetz</snm><fnm>PG</fnm></au>
    <au><snm>Pierce</snm><fnm>W</fnm></au>
    <au><snm>Maher</snm><fnm>VE</fnm></au>
    <au><snm>Zhang</snm><fnm>H</fnm></au>
    <au><snm>Tang</snm><fnm>S</fnm></au>
    <au><snm>Song</snm><fnm>P</fnm></au>
    <au><snm>Liu</snm><fnm>Q</fnm></au>
    <au><snm>Haber</snm><fnm>MT</fnm></au>
    <au><snm>Leutzinger</snm><fnm>EE</fnm></au>
    <au><snm>Al Hakim</snm><fnm>A</fnm></au>
    <au><cnm>others</cnm></au>
  </aug>
  <source>Clin. Cancer Res.</source>
  <publisher>AACR</publisher>
  <pubdate>2014</pubdate>
  <volume>20</volume>
  <issue>1</issue>
  <fpage>9</fpage>
  <lpage>-14</lpage>
</bibl>

<bibl id="B4">
  <title><p>{Targeted alpha-therapy of metastatic castration-resistant prostate
  cancer with 225Ac-PSMA-617: dosimetry estimate and empiric dose
  finding}</p></title>
  <aug>
    <au><snm>Kratochwil</snm><fnm>C</fnm></au>
    <au><snm>Bruchertseifer</snm><fnm>F</fnm></au>
    <au><snm>Rathke</snm><fnm>H</fnm></au>
    <au><snm>Bronzel</snm><fnm>M</fnm></au>
    <au><snm>Apostolidis</snm><fnm>C</fnm></au>
    <au><snm>Weichert</snm><fnm>W</fnm></au>
    <au><snm>Haberkorn</snm><fnm>U</fnm></au>
    <au><snm>Giesel</snm><fnm>FL</fnm></au>
    <au><snm>Morgenstern</snm><fnm>A</fnm></au>
  </aug>
  <source>J. Nucl. Med.</source>
  <publisher>Soc Nuclear Med</publisher>
  <pubdate>2017</pubdate>
  <volume>58</volume>
  <issue>10</issue>
  <fpage>1624</fpage>
  <lpage>-1631</lpage>
</bibl>

<bibl id="B5">
  <title><p>{Feasibility of thorium-227/radium-223 gamma-camera imaging during
  radionuclide therapy}</p></title>
  <aug>
    <au><snm>Larsson</snm><fnm>E</fnm></au>
    <au><snm>Brolin</snm><fnm>G</fnm></au>
    <au><snm>Cleton</snm><fnm>A</fnm></au>
    <au><snm>Ohlsson</snm><fnm>T</fnm></au>
    <au><snm>Lind{\'e}n</snm><fnm>O</fnm></au>
    <au><snm>Hindorf</snm><fnm>C</fnm></au>
  </aug>
  <source>Cancer Biother. Radiopharm.</source>
  <publisher>Mary Ann Liebert, Inc., publishers 140 Huguenot Street, 3rd Floor
  New~…</publisher>
  <pubdate>2020</pubdate>
  <volume>35</volume>
  <issue>7</issue>
  <fpage>540</fpage>
  <lpage>-548</lpage>
</bibl>

<bibl id="B6">
  <title><p>{Quantitative Dual-Isotope Planar Imaging of Thorium-227 and
  Radium-223 Using Defined Energy Windows}</p></title>
  <aug>
    <au><snm>Murray</snm><fnm>I</fnm></au>
    <au><snm>Rojas</snm><fnm>B</fnm></au>
    <au><snm>Gear</snm><fnm>J</fnm></au>
    <au><snm>Callister</snm><fnm>R</fnm></au>
    <au><snm>Cleton</snm><fnm>A</fnm></au>
    <au><snm>Flux</snm><fnm>GD</fnm></au>
  </aug>
  <source>Cancer Biother. Radiopharm.</source>
  <publisher>Mary Ann Liebert, Inc., publishers 140 Huguenot Street, 3rd Floor
  New~…</publisher>
  <pubdate>2020</pubdate>
  <volume>35</volume>
  <issue>7</issue>
  <fpage>530</fpage>
  <lpage>-539</lpage>
</bibl>

<bibl id="B7">
  <title><p>{Quantitative dual isotope SPECT imaging of the alpha-emitters
  Th-227 and Ra-223}</p></title>
  <aug>
    <au><snm>Ghaly</snm><fnm>M</fnm></au>
    <au><snm>Sgouros</snm><fnm>G</fnm></au>
    <au><snm>Frey</snm><fnm>E</fnm></au>
  </aug>
  <publisher>Soc Nuclear Med</publisher>
  <pubdate>2019</pubdate>
</bibl>

<bibl id="B8">
  <title><p>{Feasibility and limitations of quantitative SPECT for
  223Ra}</p></title>
  <aug>
    <au><snm>Gustafsson</snm><fnm>J</fnm></au>
    <au><snm>Rode{\~n}o</snm><fnm>E</fnm></au>
    <au><snm>M{\'\i}nguez</snm><fnm>P</fnm></au>
  </aug>
  <source>Phys. Med. Biol.</source>
  <publisher>IOP Publishing</publisher>
  <pubdate>2020</pubdate>
  <volume>65</volume>
  <issue>8</issue>
  <fpage>085012</fpage>
</bibl>

<bibl id="B9">
  <title><p>{Development of targeted alpha particle therapy for solid
  tumors}</p></title>
  <aug>
    <au><snm>Tafreshi</snm><fnm>NK</fnm></au>
    <au><snm>Doligalski</snm><fnm>ML</fnm></au>
    <au><snm>Tichacek</snm><fnm>CJ</fnm></au>
    <au><snm>Pandya</snm><fnm>DN</fnm></au>
    <au><snm>Budzevich</snm><fnm>MM</fnm></au>
    <au><snm>El Haddad</snm><fnm>G</fnm></au>
    <au><snm>Khushalani</snm><fnm>NI</fnm></au>
    <au><snm>Moros</snm><fnm>EG</fnm></au>
    <au><snm>McLaughlin</snm><fnm>ML</fnm></au>
    <au><snm>Wadas</snm><fnm>TJ</fnm></au>
    <au><cnm>others</cnm></au>
  </aug>
  <source>Molecules</source>
  <publisher>MDPI</publisher>
  <pubdate>2019</pubdate>
  <volume>24</volume>
  <issue>23</issue>
  <fpage>4314</fpage>
</bibl>

<bibl id="B10">
  <title><p>Clinical radionuclide therapy dosimetry: the quest for the “Holy
  Gray”</p></title>
  <aug>
    <au><snm>Brans</snm><fnm>B</fnm></au>
    <au><snm>Bodei</snm><fnm>L</fnm></au>
    <au><snm>Giammarile</snm><fnm>F</fnm></au>
    <au><snm>Lind{\'e}n</snm><fnm>O</fnm></au>
    <au><snm>Luster</snm><fnm>M</fnm></au>
    <au><snm>Oyen</snm><fnm>WJG</fnm></au>
    <au><snm>Tennvall</snm><fnm>J</fnm></au>
  </aug>
  <source>Eur. J. Nucl. Med. Mol. Imaging</source>
  <publisher>Springer</publisher>
  <pubdate>2007</pubdate>
  <volume>34</volume>
  <issue>5</issue>
  <fpage>772</fpage>
  <lpage>-786</lpage>
</bibl>

<bibl id="B11">
  <title><p>{223Ra-dichloride therapy of bone metastasis: optimization of SPECT
  images for quantification}</p></title>
  <aug>
    <au><snm>Benabdallah</snm><fnm>N</fnm></au>
    <au><snm>Bernardini</snm><fnm>M</fnm></au>
    <au><snm>Bianciardi</snm><fnm>M</fnm></au>
    <au><snm>Labriolle Vaylet</snm><fnm>C</fnm></au>
    <au><snm>Franck</snm><fnm>D</fnm></au>
    <au><snm>Desbr{\'e}e</snm><fnm>A</fnm></au>
  </aug>
  <source>EJNMMI research</source>
  <publisher>Springer</publisher>
  <pubdate>2019</pubdate>
  <volume>9</volume>
  <issue>1</issue>
  <fpage>1</fpage>
  <lpage>-12</lpage>
</bibl>

<bibl id="B12">
  <title><p>{SU-F-J-08: Quantitative SPECT Imaging of Ra-223 in a
  Phantom}</p></title>
  <aug>
    <au><snm>Yue</snm><fnm>J</fnm></au>
    <au><snm>Hobbs</snm><fnm>R</fnm></au>
    <au><snm>Sgouros</snm><fnm>G</fnm></au>
    <au><snm>Frey</snm><fnm>E</fnm></au>
  </aug>
  <source>Med. Phys.</source>
  <pubdate>2016</pubdate>
  <volume>43</volume>
  <issue>6</issue>
  <url>https://www.osti.gov/biblio/22632144</url>
</bibl>

<bibl id="B13">
  <title><p>{A projection-domain low-count quantitative SPECT method for
  $\alpha$-particle emitting radiopharmaceutical therapy}</p></title>
  <aug>
    <au><snm>Li</snm><fnm>Z</fnm></au>
    <au><snm>Benabdallah</snm><fnm>N</fnm></au>
    <au><snm>Abou</snm><fnm>DS</fnm></au>
    <au><snm>Baumann</snm><fnm>BC</fnm></au>
    <au><snm>Dehdashti</snm><fnm>F</fnm></au>
    <au><snm>Ballard</snm><fnm>DH</fnm></au>
    <au><snm>Liu</snm><fnm>J</fnm></au>
    <au><snm>Jammalamadaka</snm><fnm>U</fnm></au>
    <au><snm>Laforest</snm><fnm>R</fnm></au>
    <au><snm>Wahl</snm><fnm>RL</fnm></au>
    <au><snm>Thorek</snm><fnm>DLJ</fnm></au>
    <au><snm>Jha</snm><fnm>AK</fnm></au>
  </aug>
  <source>IEEE Trans. Radiat. Plasma Med. Sci.</source>
  <pubdate>2022</pubdate>
  <fpage>1</fpage>
  <lpage>1</lpage>
</bibl>

<bibl id="B14">
  <title><p>{Preclinical single photon emission computed tomography of alpha
  particle-emitting radium-223}</p></title>
  <aug>
    <au><snm>Abou</snm><fnm>DS</fnm></au>
    <au><snm>Rittenbach</snm><fnm>A</fnm></au>
    <au><snm>Tomlinson</snm><fnm>RE</fnm></au>
    <au><snm>Finley</snm><fnm>PA</fnm></au>
    <au><snm>Tsui</snm><fnm>B</fnm></au>
    <au><snm>Simons</snm><fnm>BW</fnm></au>
    <au><snm>Jha</snm><fnm>AK</fnm></au>
    <au><snm>Ulmert</snm><fnm>D</fnm></au>
    <au><snm>Riddle</snm><fnm>RC</fnm></au>
    <au><snm>Thorek</snm><fnm>DL</fnm></au>
  </aug>
  <source>Cancer Biother. Radiopharm.</source>
  <publisher>Mary Ann Liebert, Inc., publishers 140 Huguenot Street, 3rd Floor
  New~…</publisher>
  <pubdate>2020</pubdate>
  <volume>35</volume>
  <issue>7</issue>
  <fpage>520</fpage>
  <lpage>-529</lpage>
</bibl>

<bibl id="B15">
  <title><p>{Ra-223 SPECT for semi-quantitative analysis in comparison with
  Tc-99m HMDP SPECT: phantom study and initial clinical experience}</p></title>
  <aug>
    <au><snm>Owaki</snm><fnm>Y</fnm></au>
    <au><snm>Nakahara</snm><fnm>T</fnm></au>
    <au><snm>Kosaka</snm><fnm>T</fnm></au>
    <au><snm>Fukada</snm><fnm>J</fnm></au>
    <au><snm>Kumabe</snm><fnm>A</fnm></au>
    <au><snm>Ichimura</snm><fnm>A</fnm></au>
    <au><snm>Murakami</snm><fnm>M</fnm></au>
    <au><snm>Nakajima</snm><fnm>K</fnm></au>
    <au><snm>Fukushi</snm><fnm>M</fnm></au>
    <au><snm>Inoue</snm><fnm>K</fnm></au>
    <au><cnm>others</cnm></au>
  </aug>
  <source>EJNMMI research</source>
  <publisher>Springer</publisher>
  <pubdate>2017</pubdate>
  <volume>7</volume>
  <issue>1</issue>
  <fpage>1</fpage>
  <lpage>-11</lpage>
</bibl>

<bibl id="B16">
  <title><p>{Practical considerations for quantitative clinical SPECT/CT
  imaging of alpha particle emitting radioisotopes}</p></title>
  <aug>
    <au><snm>Benabdallah</snm><fnm>N</fnm></au>
    <au><snm>Scheve</snm><fnm>W</fnm></au>
    <au><snm>Dunn</snm><fnm>N</fnm></au>
    <au><snm>Silvestros</snm><fnm>D</fnm></au>
    <au><snm>Schelker</snm><fnm>P</fnm></au>
    <au><snm>Abou</snm><fnm>D</fnm></au>
    <au><snm>Jammalamadaka</snm><fnm>U</fnm></au>
    <au><snm>Laforest</snm><fnm>R</fnm></au>
    <au><snm>Li</snm><fnm>Z</fnm></au>
    <au><snm>Liu</snm><fnm>J</fnm></au>
    <au><cnm>others</cnm></au>
  </aug>
  <source>Theranostics</source>
  <publisher>Ivyspring International Publisher</publisher>
  <pubdate>2021</pubdate>
  <volume>11</volume>
  <issue>20</issue>
  <fpage>9721</fpage>
</bibl>

<bibl id="B17">
  <title><p>{Algorithms and analyses for joint spectral image reconstruction in
  Y-90 bremsstrahlung SPECT}</p></title>
  <aug>
    <au><snm>Chun</snm><fnm>SY</fnm></au>
    <au><snm>Nguyen</snm><fnm>MP</fnm></au>
    <au><snm>Phan</snm><fnm>TQ</fnm></au>
    <au><snm>Kim</snm><fnm>H</fnm></au>
    <au><snm>Fessler</snm><fnm>JA</fnm></au>
    <au><snm>Dewaraja</snm><fnm>YK</fnm></au>
  </aug>
  <source>IEEE Trans. Med. Imaging</source>
  <publisher>IEEE</publisher>
  <pubdate>2019</pubdate>
  <volume>39</volume>
  <issue>5</issue>
  <fpage>1369</fpage>
  <lpage>-1379</lpage>
</bibl>

<bibl id="B18">
  <title><p>{An SVD investigation of modeling scatter in multiple energy
  windows for improved SPECT images}</p></title>
  <aug>
    <au><snm>Kadrmas</snm><fnm>DJ</fnm></au>
    <au><snm>Frey</snm><fnm>EC</fnm></au>
    <au><snm>Tsui</snm><fnm>BM</fnm></au>
  </aug>
  <source>IEEE Trans. Nucl. Sci.</source>
  <publisher>IEEE</publisher>
  <pubdate>1996</pubdate>
  <volume>43</volume>
  <issue>4</issue>
  <fpage>2275</fpage>
  <lpage>-2284</lpage>
</bibl>

<bibl id="B19">
  <title><p>Analysis of the reconstructibility and noise properties of
  scattered photons in {T}c-99m {SPECT}</p></title>
  <aug>
    <au><snm>Kadrmas</snm><fnm>DJ</fnm></au>
    <au><snm>Frey</snm><fnm>EC</fnm></au>
    <au><snm>Tsui</snm><fnm>BM</fnm></au>
  </aug>
  <source>Phys. Med. Biol.</source>
  <publisher>NIH Public Access</publisher>
  <pubdate>1997</pubdate>
  <volume>42</volume>
  <issue>12</issue>
  <fpage>2493</fpage>
</bibl>

<bibl id="B20">
  <title><p>Quantifying the loss of information from binning list-mode
  data</p></title>
  <aug>
    <au><snm>Clarkson</snm><fnm>E</fnm></au>
    <au><snm>Kupinski</snm><fnm>M</fnm></au>
  </aug>
  <source>JOSA A</source>
  <publisher>Optical Society of America</publisher>
  <pubdate>2020</pubdate>
  <volume>37</volume>
  <issue>3</issue>
  <fpage>450</fpage>
  <lpage>-457</lpage>
</bibl>

<bibl id="B21">
  <title><p>Estimating {ROI} activity concentration with photon-processing and
  photon-counting {SPECT} imaging systems</p></title>
  <aug>
    <au><snm>Jha</snm><fnm>AK</fnm></au>
    <au><snm>Frey</snm><fnm>EC</fnm></au>
  </aug>
  <source>Proc. SPIE Med. Imag.</source>
  <pubdate>2015</pubdate>
  <volume>9412</volume>
  <fpage>94120R</fpage>
</bibl>

<bibl id="B22">
  <title><p>Singular value decomposition for photon-processing nuclear imaging
  systems and applications for reconstruction and computing null
  functions</p></title>
  <aug>
    <au><snm>Jha</snm><fnm>AK</fnm></au>
    <au><snm>Barrett</snm><fnm>HH</fnm></au>
    <au><snm>Frey</snm><fnm>EC</fnm></au>
    <au><snm>Clarkson</snm><fnm>E</fnm></au>
    <au><snm>Caucci</snm><fnm>L</fnm></au>
    <au><snm>Kupinski</snm><fnm>MA</fnm></au>
  </aug>
  <source>Phys. Med. Biol.</source>
  <publisher>IOP Publishing</publisher>
  <pubdate>2015</pubdate>
  <volume>60</volume>
  <issue>18</issue>
  <fpage>7359</fpage>
</bibl>

<bibl id="B23">
  <title><p>Towards continuous-to-continuous 3{D} imaging in the real
  world</p></title>
  <aug>
    <au><snm>Caucci</snm><fnm>L</fnm></au>
    <au><snm>Liu</snm><fnm>Z</fnm></au>
    <au><snm>Jha</snm><fnm>AK</fnm></au>
    <au><snm>Han</snm><fnm>H</fnm></au>
    <au><snm>Furenlid</snm><fnm>LR</fnm></au>
    <au><snm>Barrett</snm><fnm>HH</fnm></au>
  </aug>
  <source>Phys. Med. Biol.</source>
  <publisher>IOP Publishing</publisher>
  <pubdate>2019</pubdate>
  <volume>64</volume>
  <issue>18</issue>
  <fpage>185007</fpage>
</bibl>

<bibl id="B24">
  <title><p>A List-Mode {OSEM}-Based Attenuation and Scatter Compensation
  Method for {SPECT}</p></title>
  <aug>
    <au><snm>{Rahman}</snm><fnm>M. A.</fnm></au>
    <au><snm>{Laforest}</snm><fnm>R.</fnm></au>
    <au><snm>{Jha}</snm><fnm>A. K.</fnm></au>
  </aug>
  <source>2020 IEEE 17th International Symposium on Biomedical Imaging
  (ISBI)</source>
  <pubdate>2020</pubdate>
  <fpage>646</fpage>
  <lpage>650</lpage>
</bibl>

<bibl id="B25">
  <title><p>Joint activity/attenuation reconstruction in SPECT using photopeak
  and scatter sinograms</p></title>
  <aug>
    <au><snm>Bousse</snm><fnm>A</fnm></au>
    <au><snm>Sidlesky</snm><fnm>A</fnm></au>
    <au><snm>Roth</snm><fnm>N</fnm></au>
    <au><snm>Rashidnasab</snm><fnm>A</fnm></au>
    <au><snm>Thielemans</snm><fnm>K</fnm></au>
    <au><snm>Hutton</snm><fnm>BF</fnm></au>
  </aug>
  <source>2016 IEEE Nuclear Science Symposium, Medical Imaging Conference and
  Room-Temperature Semiconductor Detector Workshop (NSS/MIC/RTSD)</source>
  <pubdate>2016</pubdate>
  <fpage>1</fpage>
  <lpage>-4</lpage>
</bibl>

<bibl id="B26">
  <title><p>(An overview of) Synergistic reconstruction for
  multimodality/multichannel imaging methods</p></title>
  <aug>
    <au><snm>Arridge</snm><fnm>SR</fnm></au>
    <au><snm>Ehrhardt</snm><fnm>MJ</fnm></au>
    <au><snm>Thielemans</snm><fnm>K</fnm></au>
  </aug>
  <source>Philosophical Transactions of the Royal Society A</source>
  <publisher>The Royal Society Publishing</publisher>
  <pubdate>2021</pubdate>
  <volume>379</volume>
  <issue>2200</issue>
  <fpage>20200205</fpage>
</bibl>

<bibl id="B27">
  <title><p>Fisher information analysis of list-mode SPECT emission data for
  joint estimation of activity and attenuation distribution</p></title>
  <aug>
    <au><snm>Rahman</snm><fnm>MA</fnm></au>
    <au><snm>Zhu</snm><fnm>Y</fnm></au>
    <au><snm>Clarkson</snm><fnm>E</fnm></au>
    <au><snm>Kupinski</snm><fnm>MA</fnm></au>
    <au><snm>Frey</snm><fnm>EC</fnm></au>
    <au><snm>Jha</snm><fnm>AK</fnm></au>
  </aug>
  <source>Inverse Problems</source>
  <publisher>IOP Publishing</publisher>
  <pubdate>2020</pubdate>
  <volume>36</volume>
  <issue>8</issue>
  <fpage>084002</fpage>
</bibl>

<bibl id="B28">
  <title><p>Novel scatter compensation of list-mode {PET} data using spatial
  and energy dependent corrections</p></title>
  <aug>
    <au><snm>Gu{\'e}rin</snm><fnm>B</fnm></au>
    <au><snm>El Fakhri</snm><fnm>G</fnm></au>
  </aug>
  <source>IEEE Trans. Med. Imaging</source>
  <publisher>IEEE</publisher>
  <pubdate>2010</pubdate>
  <volume>30</volume>
  <issue>3</issue>
  <fpage>759</fpage>
  <lpage>-773</lpage>
</bibl>

<bibl id="B29">
  <title><p>Retrieving information from scattered photons in medical
  imaging</p></title>
  <aug>
    <au><snm>Jha</snm><fnm>AK</fnm></au>
  </aug>
  <publisher>The University of Arizona.</publisher>
  <pubdate>2013</pubdate>
</bibl>

<bibl id="B30">
  <title><p>Joint reconstruction of activity and attenuation map using {LM}
  {SPECT} emission data</p></title>
  <aug>
    <au><snm>Jha</snm><fnm>AK</fnm></au>
    <au><snm>Clarkson</snm><fnm>E</fnm></au>
    <au><snm>Kupinski</snm><fnm>MA</fnm></au>
    <au><snm>Barrett</snm><fnm>HH</fnm></au>
  </aug>
  <source>Proc. SPIE Med. Imag.</source>
  <pubdate>2013</pubdate>
  <volume>8668</volume>
  <fpage>86681W</fpage>
</bibl>

<bibl id="B31">
  <title><p>{List-mode likelihood}</p></title>
  <aug>
    <au><snm>Barrett</snm><fnm>HH</fnm></au>
    <au><snm>White</snm><fnm>T</fnm></au>
    <au><snm>Parra</snm><fnm>LC</fnm></au>
  </aug>
  <source>JOSA A</source>
  <publisher>Optica Publishing Group</publisher>
  <pubdate>1997</pubdate>
  <volume>14</volume>
  <issue>11</issue>
  <fpage>2914</fpage>
  <lpage>-2923</lpage>
</bibl>

<bibl id="B32">
  <title><p>{{\"U}ber die Streuung von Strahlung durch freie Elektronen nach
  der neuen relativistischen Quantendynamik von Dirac}</p></title>
  <aug>
    <au><snm>Klein</snm><fnm>O</fnm></au>
    <au><snm>Nishina</snm><fnm>Y</fnm></au>
  </aug>
  <source>Zeitschrift f{\"u}r Physik</source>
  <publisher>Springer</publisher>
  <pubdate>1929</pubdate>
  <volume>52</volume>
  <issue>11</issue>
  <fpage>853</fpage>
  <lpage>-868</lpage>
</bibl>

<bibl id="B33">
  <title><p>{Maximum likelihood reconstruction for emission
  tomography}</p></title>
  <aug>
    <au><snm>Shepp</snm><fnm>LA</fnm></au>
    <au><snm>Vardi</snm><fnm>Y</fnm></au>
  </aug>
  <source>IEEE Trans. Med. Imaging</source>
  <publisher>IEEE</publisher>
  <pubdate>1982</pubdate>
  <volume>1</volume>
  <issue>2</issue>
  <fpage>113</fpage>
  <lpage>-122</lpage>
</bibl>

<bibl id="B34">
  <title><p>EM reconstruction algorithms for emission and transmission
  tomography</p></title>
  <aug>
    <au><snm>Lange</snm><fnm>K</fnm></au>
    <au><snm>Carson</snm><fnm>R</fnm></au>
    <au><cnm>others</cnm></au>
  </aug>
  <source>J. Comput. Assist. Tomogr.</source>
  <pubdate>1984</pubdate>
  <volume>8</volume>
  <issue>2</issue>
  <fpage>306</fpage>
  <lpage>-16</lpage>
</bibl>

<bibl id="B35">
  <title><p>{List-mode likelihood: EM algorithm and image quality estimation
  demonstrated on 2-D PET}</p></title>
  <aug>
    <au><snm>Parra</snm><fnm>L</fnm></au>
    <au><snm>Barrett</snm><fnm>HH</fnm></au>
  </aug>
  <source>IEEE Trans. Med. Imaging</source>
  <publisher>IEEE</publisher>
  <pubdate>1998</pubdate>
  <volume>17</volume>
  <issue>2</issue>
  <fpage>228</fpage>
  <lpage>-235</lpage>
</bibl>

<bibl id="B36">
  <title><p>{A globally convergent regularized ordered-subset EM algorithm for
  list-mode reconstruction}</p></title>
  <aug>
    <au><snm>Khurd</snm><fnm>P</fnm></au>
    <au><snm>Hsiao</snm><fnm>T</fnm></au>
    <au><snm>Rangarajan</snm><fnm>A</fnm></au>
    <au><snm>Gindi</snm><fnm>G</fnm></au>
  </aug>
  <source>IEEE Trans. Nucl. Sci.</source>
  <publisher>IEEE</publisher>
  <pubdate>2004</pubdate>
  <volume>51</volume>
  <issue>3</issue>
  <fpage>719</fpage>
  <lpage>-725</lpage>
</bibl>

<bibl id="B37">
  <title><p>{Accelerated image reconstruction using ordered subsets of
  projection data}</p></title>
  <aug>
    <au><snm>Hudson</snm><fnm>HM</fnm></au>
    <au><snm>Larkin</snm><fnm>RS</fnm></au>
  </aug>
  <source>IEEE Trans. Med. Imaging</source>
  <publisher>IEEE</publisher>
  <pubdate>1994</pubdate>
  <volume>13</volume>
  <issue>4</issue>
  <fpage>601</fpage>
  <lpage>-609</lpage>
</bibl>

<bibl id="B38">
  <title><p>Statistical texture synthesis of mammographic images with clustered
  lumpy backgrounds</p></title>
  <aug>
    <au><snm>Bochud</snm><fnm>FO</fnm></au>
    <au><snm>Abbey</snm><fnm>CK</fnm></au>
    <au><snm>Eckstein</snm><fnm>MP</fnm></au>
  </aug>
  <source>Opt. Express</source>
  <publisher>Optical Society of America</publisher>
  <pubdate>1999</pubdate>
  <volume>4</volume>
  <issue>1</issue>
  <fpage>33</fpage>
  <lpage>-43</lpage>
</bibl>

<bibl id="B39">
  <title><p>{Quantitative imaging of 223Ra-chloride (Alpharadin) for targeted
  alpha-emitting radionuclide therapy of bone metastases}</p></title>
  <aug>
    <au><snm>Hindorf</snm><fnm>C</fnm></au>
    <au><snm>Chittenden</snm><fnm>S</fnm></au>
    <au><snm>Aksnes</snm><fnm>AK</fnm></au>
    <au><snm>Parker</snm><fnm>C</fnm></au>
    <au><snm>Flux</snm><fnm>GD</fnm></au>
  </aug>
  <source>Nucl. Med. Commun.</source>
  <publisher>LWW</publisher>
  <pubdate>2012</pubdate>
  <volume>33</volume>
  <issue>7</issue>
  <fpage>726</fpage>
  <lpage>-732</lpage>
</bibl>

<bibl id="B40">
  <title><p>Medical imaging signals and systems</p></title>
  <aug>
    <au><snm>Prince</snm><fnm>JL</fnm></au>
    <au><snm>Links</snm><fnm>JM</fnm></au>
  </aug>
  <publisher>Pearson Prentice Hall Upper Saddle River</publisher>
  <pubdate>2006</pubdate>
  <volume>37</volume>
</bibl>

<bibl id="B41">
  <title><p>{Does OSEM achieve the lowest variance?}</p></title>
  <aug>
    <au><snm>Cloquet</snm><fnm>C</fnm></au>
    <au><snm>Defrise</snm><fnm>M</fnm></au>
  </aug>
  <source>2011 IEEE Nuclear Science Symposium Conference Record</source>
  <pubdate>2011</pubdate>
  <fpage>2360</fpage>
  <lpage>-2365</lpage>
</bibl>

<bibl id="B42">
  <title><p>{Activity determination of 227Ac and 223Ra by means of liquid
  scintillation counting and determination of nuclear decay data}</p></title>
  <aug>
    <au><snm>Kossert</snm><fnm>K</fnm></au>
    <au><snm>Bokeloh</snm><fnm>K</fnm></au>
    <au><snm>Dersch</snm><fnm>R</fnm></au>
    <au><snm>N{\"a}hle</snm><fnm>O</fnm></au>
  </aug>
  <source>Appl. Radiat. Isot.</source>
  <publisher>Elsevier</publisher>
  <pubdate>2015</pubdate>
  <volume>95</volume>
  <fpage>143</fpage>
  <lpage>-152</lpage>
</bibl>

<bibl id="B43">
  <title><p>{Determination of photon emission probability for the main gamma
  ray and half-life measurements of 64Cu}</p></title>
  <aug>
    <au><snm>Pibida</snm><fnm>L</fnm></au>
    <au><snm>Zimmerman</snm><fnm>B</fnm></au>
    <au><snm>Bergeron</snm><fnm>DE</fnm></au>
    <au><snm>Fitzgerald</snm><fnm>R</fnm></au>
    <au><snm>Cessna</snm><fnm>JT</fnm></au>
    <au><snm>King</snm><fnm>L</fnm></au>
  </aug>
  <source>Appl. Radiat. Isot.</source>
  <publisher>Elsevier</publisher>
  <pubdate>2017</pubdate>
  <volume>129</volume>
  <fpage>6</fpage>
  <lpage>-12</lpage>
</bibl>

<bibl id="B44">
  <title><p>{Precise measurements of the absolute $\gamma$-ray emission
  probabilities of 223Ra and decay progeny in equilibrium}</p></title>
  <aug>
    <au><snm>Collins</snm><fnm>SM</fnm></au>
    <au><snm>Pearce</snm><fnm>AK</fnm></au>
    <au><snm>Regan</snm><fnm>PH</fnm></au>
    <au><snm>Keightley</snm><fnm>JD</fnm></au>
  </aug>
  <source>Appl. Radiat. Isot.</source>
  <publisher>Elsevier</publisher>
  <pubdate>2015</pubdate>
  <volume>102</volume>
  <fpage>15</fpage>
  <lpage>-28</lpage>
</bibl>

<bibl id="B45">
  <title><p>{A model for multiple scatters in fully 3D PET}</p></title>
  <aug>
    <au><snm>Goggin</snm><fnm>AS</fnm></au>
    <au><snm>Ollinger</snm><fnm>JM</fnm></au>
  </aug>
  <source>Proceedings of 1994 IEEE Nuclear Science Symposium-NSS'94</source>
  <pubdate>1994</pubdate>
  <volume>4</volume>
  <fpage>1609</fpage>
  <lpage>-1613</lpage>
</bibl>

<bibl id="B46">
  <title><p>{Model-based scatter correction for fully 3D PET}</p></title>
  <aug>
    <au><snm>Ollinger</snm><fnm>JM</fnm></au>
  </aug>
  <source>Phys. Med. Biol.</source>
  <publisher>IOP Publishing</publisher>
  <pubdate>1996</pubdate>
  <volume>41</volume>
  <issue>1</issue>
  <fpage>153</fpage>
</bibl>

<bibl id="B47">
  <title><p>{A new method for modeling the spatially-variant, object-dependent
  scatter response function in SPECT}</p></title>
  <aug>
    <au><snm>Frey</snm><fnm>EC</fnm></au>
    <au><snm>Tsui</snm><fnm>BMW</fnm></au>
  </aug>
  <source>1996 IEEE Nuclear Science Symposium. Conference Record</source>
  <pubdate>1996</pubdate>
  <volume>2</volume>
  <fpage>1082</fpage>
  <lpage>-1086</lpage>
</bibl>

</refgrp>
} 




\section*{Figures}
\begin{figure}[h!]
\centering
\includegraphics[height = 3in]{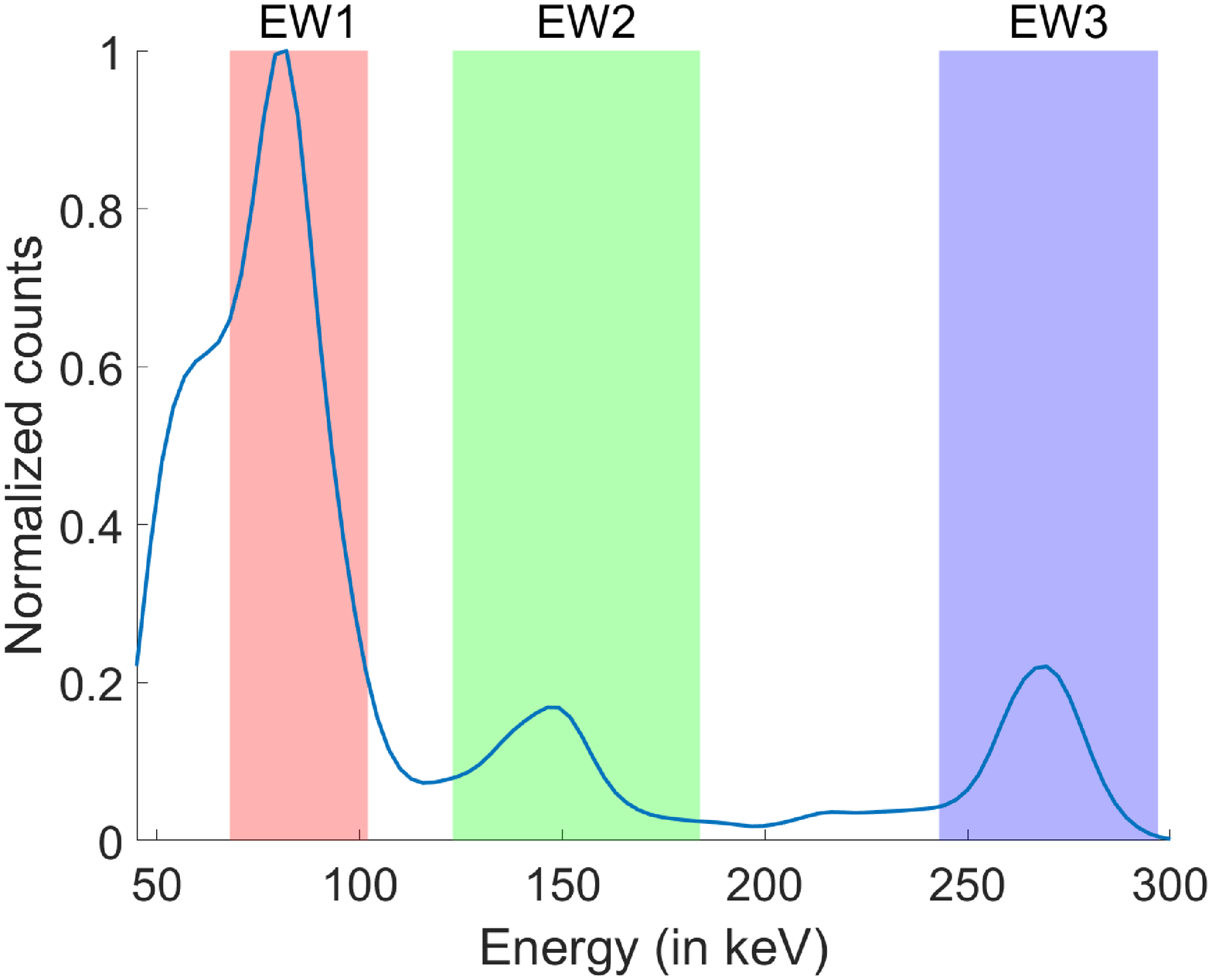}
\caption{A Monte Carlo-simulated energy spectrum for $\ra_223$-based $\alpha$-RPT SPECT in 2-D. The activity map was a single-voxel 2-D phantom and the attenuation map considered is shown in Fig.~\ref{fig:synth_true_act_attn}b. The Monte Carlo simulation process is as described in Sec.~\ref{sec:agreement} except for including multiple order of scatters. The three photopeak energy windows are shown in color. (EW1 = 68-102 keV, EW2 = 123-184 keV, EW3= 243-297 keV)}
\label{fig:MC_spectra}
\end{figure}

\begin{figure}[h!]
\centering
\includegraphics[height = 4.5in]{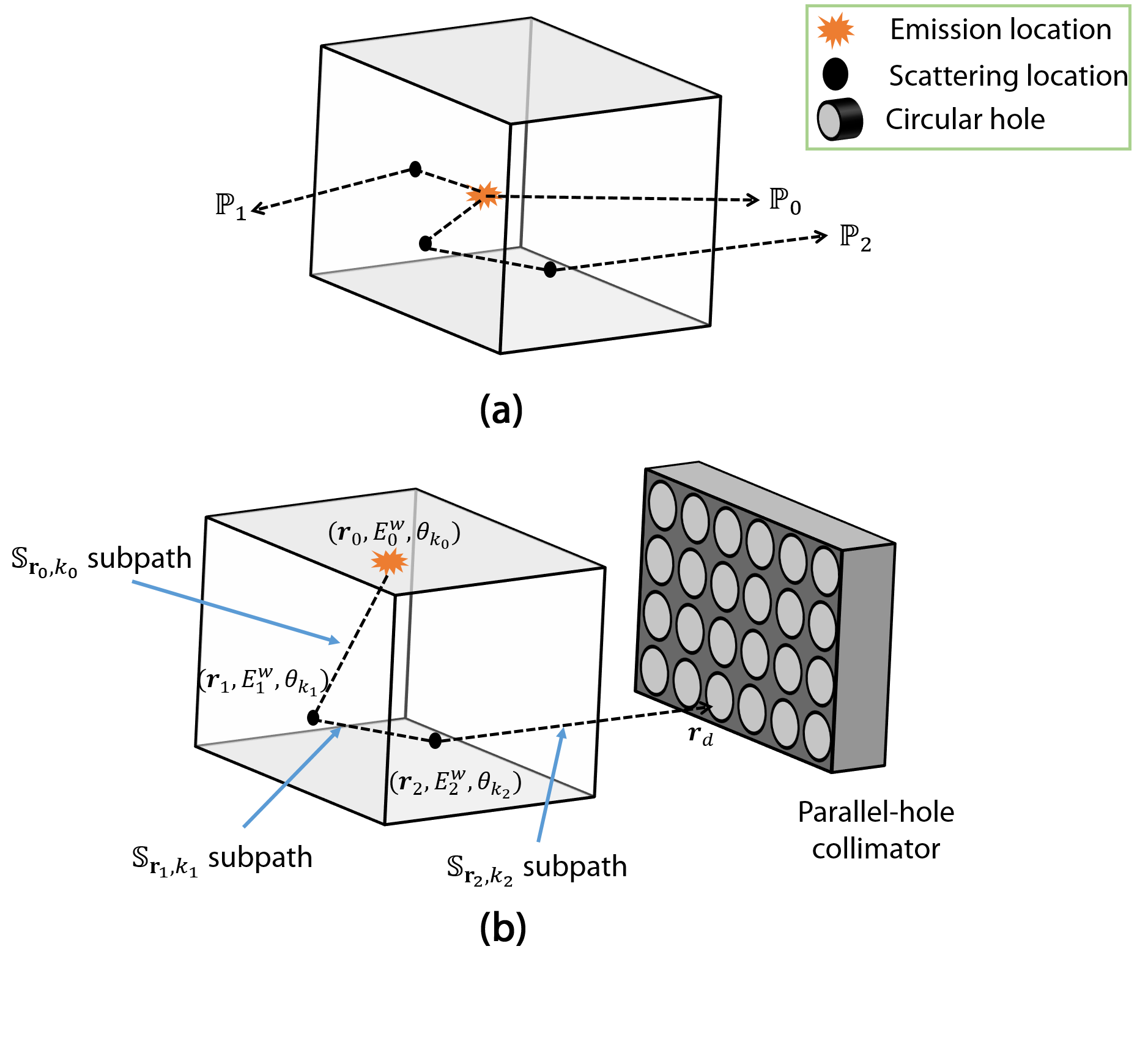}
\caption{(a) A schematic illustrating the concept of path. Here, $\P_0$ denotes a path where emitted photon is not scattered or absorped within the field of view. $\P_1$ denotes a path where emitted photon is scattered once and $\P_2$ denotes a path where emitted photon is scattered twice. (b) A schematic describing the various notations used to describe a path. At each emission or scattering location, $(r_m,E_m^w,\theta_{k_m})$ denotes the location of emission or scatter, energy at the emission or after scatter and the outgoing direction of the photon, respectively. The notation of each subpath is also demonstrated in the schematic.}
\label{fig:spect_system}
\end{figure}

\begin{figure}[h!]
\centering
\includegraphics[height = 1.6in]{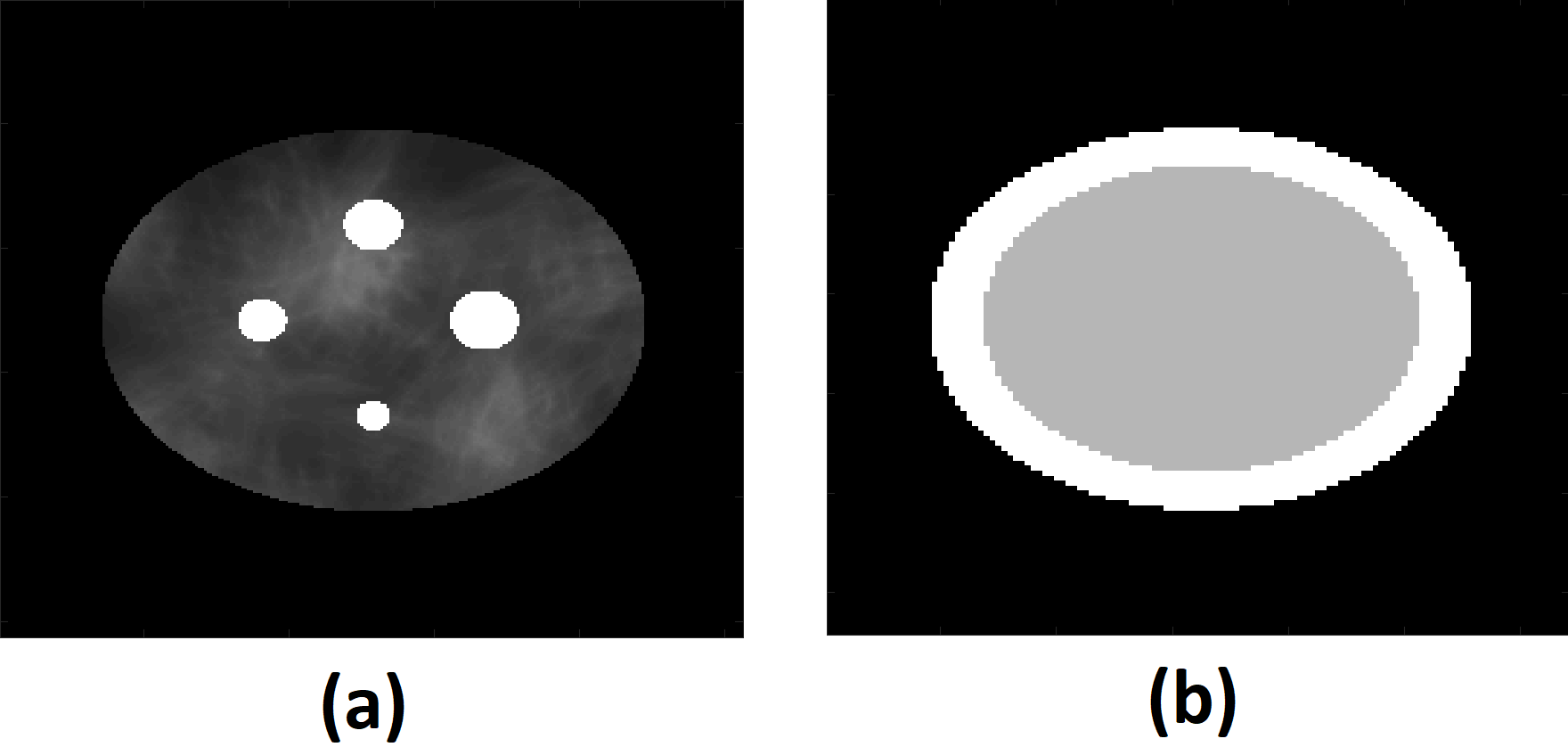}
\caption{For the synthetic phantom, (a) a realization of the activity distribution and (b) the attenuation distribution. }
\label{fig:synth_true_act_attn}
\end{figure}

\begin{figure}[h!]
\centering
\includegraphics[height = 2in]{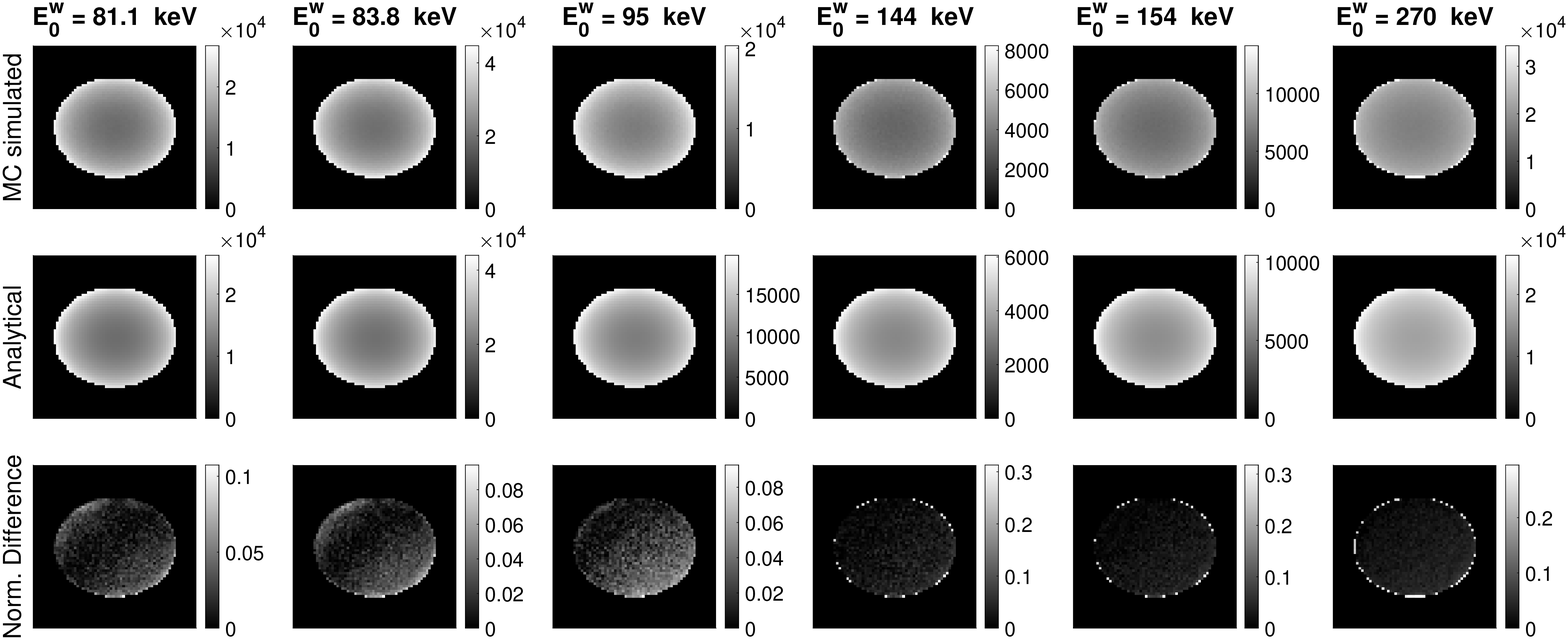}
\caption{Sensitivity images generated from MC, path-based modeling (Analytical) and normalized difference between these two methods. Images in each column indicate a fixed emitted energy value. The energy values are 81.1, 83.8, 95, 144, 154 and 270 keV in order. To calculate the sensitivity image, we consider all the energy windows (EW1+EW2+EW3(PP)). The normalized difference map denotes the absolute error in counts normalized by the counts in MC simulation. Thus, the map signifies the relative difference with reference to MC simulated sensitivity map.}
\label{fig:mc_valid}
\end{figure}

\begin{figure}[h!]
\centering
\includegraphics[height = 1.3in]{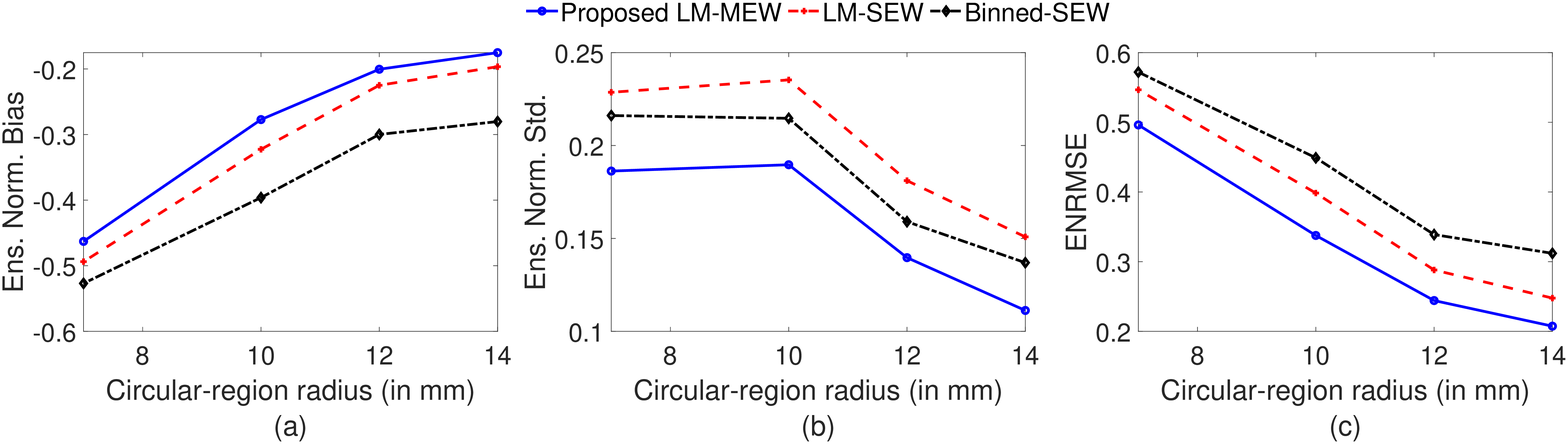}
\label{fig:comp_nrmse}
\caption{(a) Ensemble normalized bias, (b) ensemble normalized standard deviation and (c) ENRMSE as a function of circular-region radius for proposed LM-MEW method, LM-SEW method and Binned-SEW method.}
\label{fig:comp_all}
\end{figure}

\begin{figure}[h!]
\centering
\includegraphics[height = 1.3in]{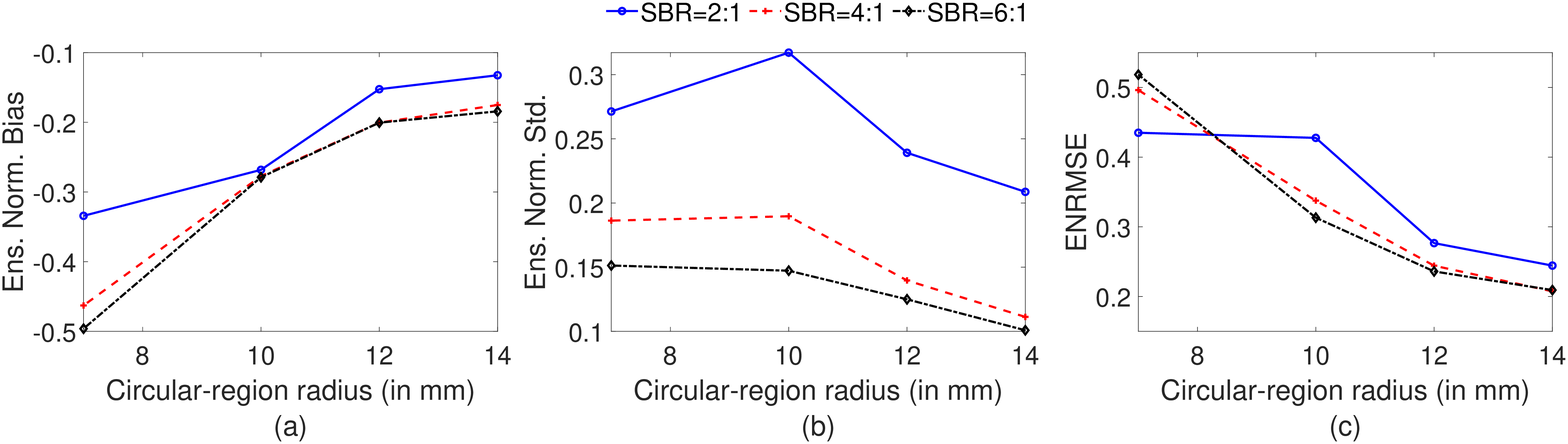}
\label{fig:diftum_nrmse}
\caption{(a) Ensemble normalized bias, (b) ensemble normalized standard deviation and (c) ENRMSE as a function of circular-region radius for different SBR values with the proposed method.}
\label{fig:diftum}
\end{figure}

\begin{figure}[h!]
\centering
\includegraphics[height = 2.8in]{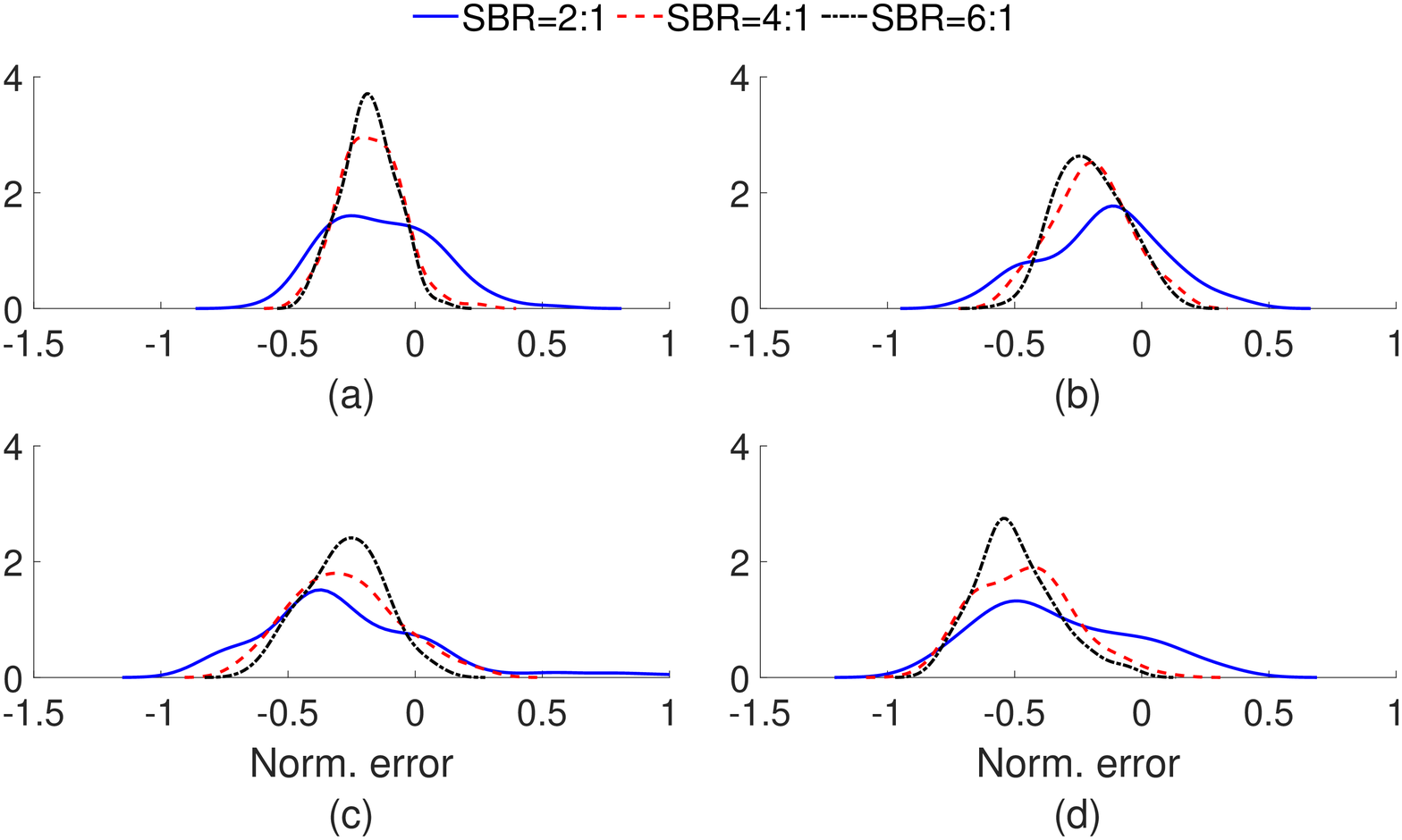}
\caption{The distribution of normalized error in the estimated uptake as a function of SBR values for the circular-region radius of (a) 14 mm, (b) 12 mm, (c) 10 mm and (d) 7 mm. The normalized error for the $k^{th}$ circular region, $n^{th}$ noise realization and $s^{th}$ object realization was computed by: $\dfrac{\hat{y}_{sk}^{n}-y_{sk}}{y_{sk}}$. The distributions were generated by kernel estimator.}
\label{fig:hist_error}
\end{figure}



  
\end{backmatter}
\end{document}